\documentclass[twocolumn]{aa}  

\usepackage{graphicx}
\usepackage{txfonts}
\usepackage{xcolor}

\usepackage[colorlinks=true,urlcolor=blue,citecolor=blue,linkcolor=blue]{hyperref}
\newcommand{\bifrost}{{\fontfamily{lmtt}\selectfont Bifrost}}
\newcommand{\simulation}{{\fontfamily{lmtt}\selectfont ch024031$\_$by200bz005}}

\begin{document} 

    \title{Stirring the Base of the Solar Wind: On Heat Transfer and Vortex Formation}
   
   \titlerunning{Stirring the Base of the Solar Wind}
   \authorrunning{Finley et al.}

   \author{A. J. Finley\inst{1}
   %\author[0000-0002-3020-9409]{Adam J.~Finley}
          \and
          A. S. Brun\inst{1}
          %\author[0000-0002-1729-8267]{Allan Sacha Brun}
          \and 
          M. Carlsson\inst{2}\inst{3}
          %\author[0000-0001-9218-3139]{Mats Carlsson}
          \and 
          M. Szydlarski\inst{2}\inst{3}
          %\author[0000-0002-9115-4448]{Mikolaj Szydlarski}
          \and 
          V. Hansteen\inst{2}\inst{3}\inst{4}\inst{5}
          %\author[0000-0003-0975-6659]{Viggo Hansteen}
          \and 
          M. Shoda\inst{6}
          %\author[??]{Munehito Shoda}
          }

    \institute{Department of Astrophysics-AIM, University of Paris-Saclay and University of Paris, CEA, CNRS, Gif-sur-Yvette Cedex 91191, France \\ \email{adam.finley@cea.fr} \and
        Rosseland Centre for Solar Physics, University of Oslo, P.O. Box 1029 Blindern, NO-0315 Oslo, Norway \\ \and
        Institute of Theoretical Astrophysics, University of Oslo, P.O. Box 1029, Blindern, NO-0315 Oslo, Norway \\ \and
        Bay Area Environmental Research Institute, Moffett Field, CA 94035, USA \\ \and
        Lockheed Martin Solar and Astrophysics Laboratory, Palo Alto, CA 94304, USA \\ \and
        National Astronomical Observatory of Japan, National Institutes of Natural Sciences, 2-21-1 Osawa, Mitaka, Tokyo, 181-8588, Japan }

   \date{Received March 30, 2022; accepted - -, -}

% \abstract{}{}{}{}{} 
% 5 {} token are mandatory
 
  \abstract{Current models of the solar wind must approximate (or ignore) the small-scale dynamics within the solar atmosphere, however these are likely important in shaping the emerging wave-turbulence spectrum and ultimately heating/accelerating the coronal plasma.}{This study strives to make connections between small-scale vortex motions at the base of the solar wind and the resulting heating/acceleration of coronal plasma.}{The \bifrost{} code produces realistic simulations of the solar atmosphere that facilitate the analysis of spatial and temporal scales which are currently at, or beyond, the limit of modern solar telescopes. For this study, the \bifrost{} simulation is configured to represent the solar atmosphere in a coronal hole region, from which the fast solar wind emerges. The simulation extends from the upper-convection zone (2.5 Mm below the photosphere) to the low-corona (14.5 Mm above the photosphere), with a horizontal extent of 24 Mm x 24 Mm. The network of magnetic funnels in the computational domain influence the movement of plasma, and the propagation of magnetohydrodynamic waves, into the low-corona.}{The twisting of the coronal magnetic field by photospheric flows, efficiently injects energy into the low-corona. Poynting fluxes of up to $2-4$ kWm$^{-2}$ are commonly observed inside twisted magnetic structures with diameters in the low-corona of 1-5 Mm. Torsional Alfv\'en waves are favourably transmitted along these structures, and will subsequently escape into the solar wind. However, reflections of these waves from the upper boundary condition make it difficult to unambiguously quantify the emerging Alfv\'en wave-energy flux. }{This study represents a first step in quantifying the conditions at the base of the solar wind using \bifrost{} simulations. It is shown that the coronal magnetic field is readily braided and twisted by photospheric flows. Temperature and density contrasts form between regions with active stirring motions and those without. Stronger whirlpool-like flows in the convection, concurrent with magnetic concentrations, launch torsional Alfv\'en waves up through the magnetic funnel network, which are expected to enhance the turbulent generation of magnetic switchbacks in the solar wind. } 

   \keywords{Solar Wind -- 
                Solar Atmosphere
                    }

   \maketitle
%
%-------------------------------------------------------------------

\section{Introduction}
The term \textit{solar wind} describes the mixture of charged particles and magnetic flux that is constantly emerging from the Sun and filling the Heliosphere \citep{parker1958dynamics}. Studies of the heating and acceleration of the solar wind have gained momentum during the past several decades \citep[][and references therein]{marsch2018solar,verscharen2019multi}, motivated (in part) by technological advances that rely heavily on space-based infrastructure\footnote{Parallels can be drawn with the 19th century, which saw many advances in Meteorology due to the increasing demand for safe travel/transportation via the oceans \citep{hearn2002tracks}.} (assets that need to be protected from extreme \textit{space weather}; \citealp{varela2021mhd}). There are many physical processes that could explain the heating and acceleration of the solar wind \citep[see reviews by][]{parnell2012contemporary,hansteen2012}, most have been studied theoretically, observed remotely, and/or measured in-situ. Yet it appears that no single mechanism alone can explain the heating of the corona \citep[see discussion in][]{klimchuk2015key}. Instead, a number of different processes are likely occurring simultaneously, and with varying degrees of significance, under the different conditions found in the solar atmosphere; from coronal holes to active regions \citep{cranmer2019properties}. 

\begin{figure*}
 \centering
  \includegraphics[trim=0cm 0cm 0cm 1cm, clip, width=0.95\textwidth]{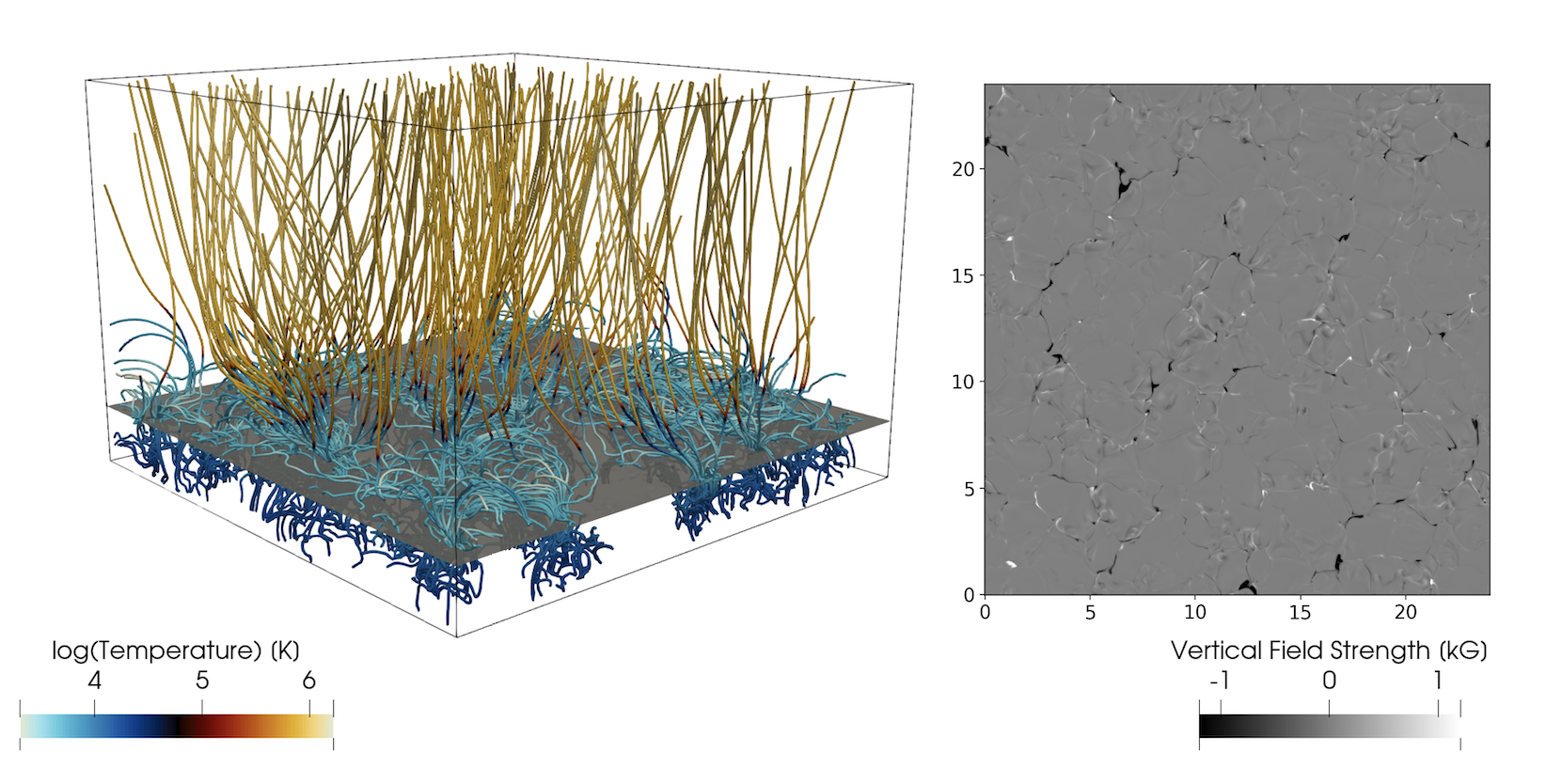}
   \caption{A 3D visualisation of \simulation{} at $t=40$~mins. The computational domain spans 24 Mm x 24 Mm x 17 Mm. Left: Magnetic field lines are coloured by temperature from blue in the upper-convection zone ($\approx$5,800 K), to light blue in the temperature minimum region ($\approx$3,200 K), and finally yellow in the low-corona ($\approx$1 MK). Right: The vertical magnetic field strength at the photosphere (Z = 0 Mm).  }
   \label{figureFieldStructure}
\end{figure*}

Despite significant advances, the solar wind remains challenging to model (and importantly forecast) due to the large range of scales that need to be incorporated. It is well known that energy is injected into the corona at all scales, from MHD waves \citep{ nutto2012modification, van2020magnetohydrodynamic} to flares and coronal mass-ejections with global extent \citep{aschwanden2017global, green2018origin, wyper2018breakout}. As well as the range of spatial scales, heating events take place on a variety of timescales \citep{hollweg1973alfvtn, viall2017survey}. Putting aside large-scale eruptions, to focus on the quasi-steady input of energy into the corona, it is possible to distinguish three broad magnetic configurations of the solar atmosphere for modelling purposes. These are the quiet sun \citep{danilovic2010probing, rempel2014numerical}, active/enhanced regions \citep{carlsson2016publicly,chen2021comprehensive}, and coronal holes \citep{wojcik2019two}. In each of these configurations, energy is channelled from the convection into the low-corona. Coronal holes are the dominant source of the solar wind in the Heliosphere \citep{cranmer2017origins, stansby2021active}, typically producing the \textit{fast solar wind} \citep{mccomas2008weaker, ebert2009bulk, macneil2020parker, wang2020small}. The magnetic field configuration of a coronal hole is relatively simple, compared with the quiet sun and active regions, given that the field is principally open to the solar wind \citep{lowder2017coronal, hofmeister2019photospheric}. However, there are still a range of dynamic processes taking place, such as the braiding of magnetic field lines \citep{wedemeyer2012magnetic, wedemeyer2013magnetic, huang2018magnetic}, and the emergence of new magnetic flux \citep{murray2009simulations}. These can trigger the formation of jets \citep{shen2017solar, yang2017formation}, and other phenomena, that are then observed as spicules \citep{martinez2017generation, bose2021spicules} or fibrils \citep{hansteen2006dynamic, leenaarts2015fibrils}. 

Vortical flows (so called \textit{solar tornadoes}) have garnered significant interest in recent years, as they can efficiently channel mass and energy from the photosphere, through the chromosphere and into the low-corona. These tornado-like events have been observed across the solar surface \citep[e.g.][]{wedemeyer2009small}. Starting in the photosphere, whirlpool-like flows form in down-flow lanes of the granulation, where strong magnetic elements can accumulate and suppress the magnetoconvection. The photospheric plasma twists the magnetic field, in turn this acts on the material above, forcing a swirling/rotation of the chromospheric plasma \citep{bonet2010sunrise, park2016first, tziotziou2018persistent} and in some cases coronal plasma \citep{wedemeyer2012magnetic}. Typically, these flows are observed to have a diameter of 0.5 - 2 Mm in the photosphere \citep[see][and references therein]{murabito2020unveiling}, though numerical models suggest the average size to be smaller \citep[less than 0.1 Mm across;][]{liu2019co}. These flows are expected to expand in the chromosphere to around 1.5 - 5 Mm \citep{battaglia2021alfv}, and may continue higher in the low-corona. In addition to vortical flows, there exist a range of linear (horizontal/vertical) drivers in the photosphere \citep[typically identified by the motion of magnetic bright points, e.g.][]{bodnarova2014dynamics}. These linear motions were at first more readily identified \citep[][]{nisenson2003motions}, however magnetohydrodynamic modelling typically shows that the emerging linear wave-energy flux is not as efficiently transmitted into the low-corona \citep{vigeesh2012three}, when compared with the torsional wave-energy flux.

Along with low-frequency Alfv\'en waves, a range of MHD waves are generated in the photosphere and travel out along the coronal magnetic field lines \citep[see review of][]{srivastava2021chromospheric}, including acoustic-gravity waves \citep{vigeesh2017internal, fleck2021acoustic}, magnetosonic waves \citep{yadav2021slow}, kink waves \citep{tiwari2019damping, morton2021weak}, and Alfv\'en waves \citep{jess2009alfven, wang2020simulation, pelekhata2021solar}. The propagation, and characteristics, of these waves are likely modified by the configuration of the overlying magnetic field \citep{bogdan2003waves,snow2018magnetic}. Subsequently these waves may generate magnetosonic shocks \citep{wang2021fast}, or otherwise undergo mode conversion \citep{schunker2006magnetic, shoda2018high}, and/or phase mixing \citep{pagano2017contribution, shoda2018anisotropic, boocock2021enhanced}, before finally escaping into the solar wind, where they may be further subject to the parametric decay instability \citep{reville2018parametric}. In most current models of the solar wind, Alfv\'en waves, generated by the shaking of the magnetic field in the photosphere, are responsible for the heating and acceleration of the solar wind when they undergo dissipation with counter-propagating (reflected) waves in the corona \citep{cranmer2005generation,van2011heating}. 

Heliospheric models tend to account for the turbulent process of Alfv\'en wave-dissipation either, on a sub-grid scale \citep{usmanov2011solar, chhiber2021large}, or with a simplified physical Wentzel-Kramer-Brillouin (WKB) description \citep{van2014alfven}. In the simplest case, one additional equation that describes the wave-energy conservation can be added to the set of standard MHD equations \citep[see][]{reville2020role}. In this case, three values must be specified \textit{a priori}: 1) input wave amplitude (or equivalently the input Alfv\'en wave-energy), 2) the length scale of dissipation, and 3) the degree to which waves are reflected as they propagate into the Heliosphere. To resolve fully the Alfv\'en wave-turbulence, requires a model with an increased spatial/temporal resolution and subsequently, to balance the computational cost, a smaller physical domain \citep{suzuki2011self, shoda2019three, shoda2020alfv}. These models reproduce many of the properties of the solar wind observed \textit{in-situ}, but they still rely on an injected spectrum of Alfv\'enic fluctuations. Generally, this spectrum is prescribed based on current knowledge of the smallest scales on the Sun \citep[see review of][]{jess2015multiwavelength}. However current observations (from the Swedish 1-m Solar Telescope, Hinode, etc) are unlikely to capture all the spatial scales and frequencies at which energy is being transported into the low-corona \citep[though the Daniel K. Inouye Solar Telescope (DKIST) will soon be fully operational;][]{rimmele2020daniel}. As some small scales remain out of reach of current instruments, it is reasonable to turn toward advanced numerical simulations to derive the conditions at the base of the solar wind.

In this study, we examine the small-scale dynamics found inside a realistic RMHD model of a coronal hole, in connection with how they may relate to the heating and acceleration of the solar wind. Section 2 presents the \bifrost{} code \citep{gudiksen2011stellar} and the computational setup that is used. Section 3 discusses the dynamical processes observed within the simulated coronal hole, including the braiding of magnetic flux and the formation of vortex-like flows and structures. In Section 4, we evaluate a number of quantities which are of interest to the solar wind modelling community, such as the amplitude of horizontal fluctuations and the strength of the vertical Poynting flux. Finally, we conclude by summarising how these results could be used to improve future solar wind modelling.

\section{Numerical Simulation}

\begin{figure}[t!]
 \centering
  \includegraphics[trim=0cm 0cm 0cm 0cm, width=0.45\textwidth]{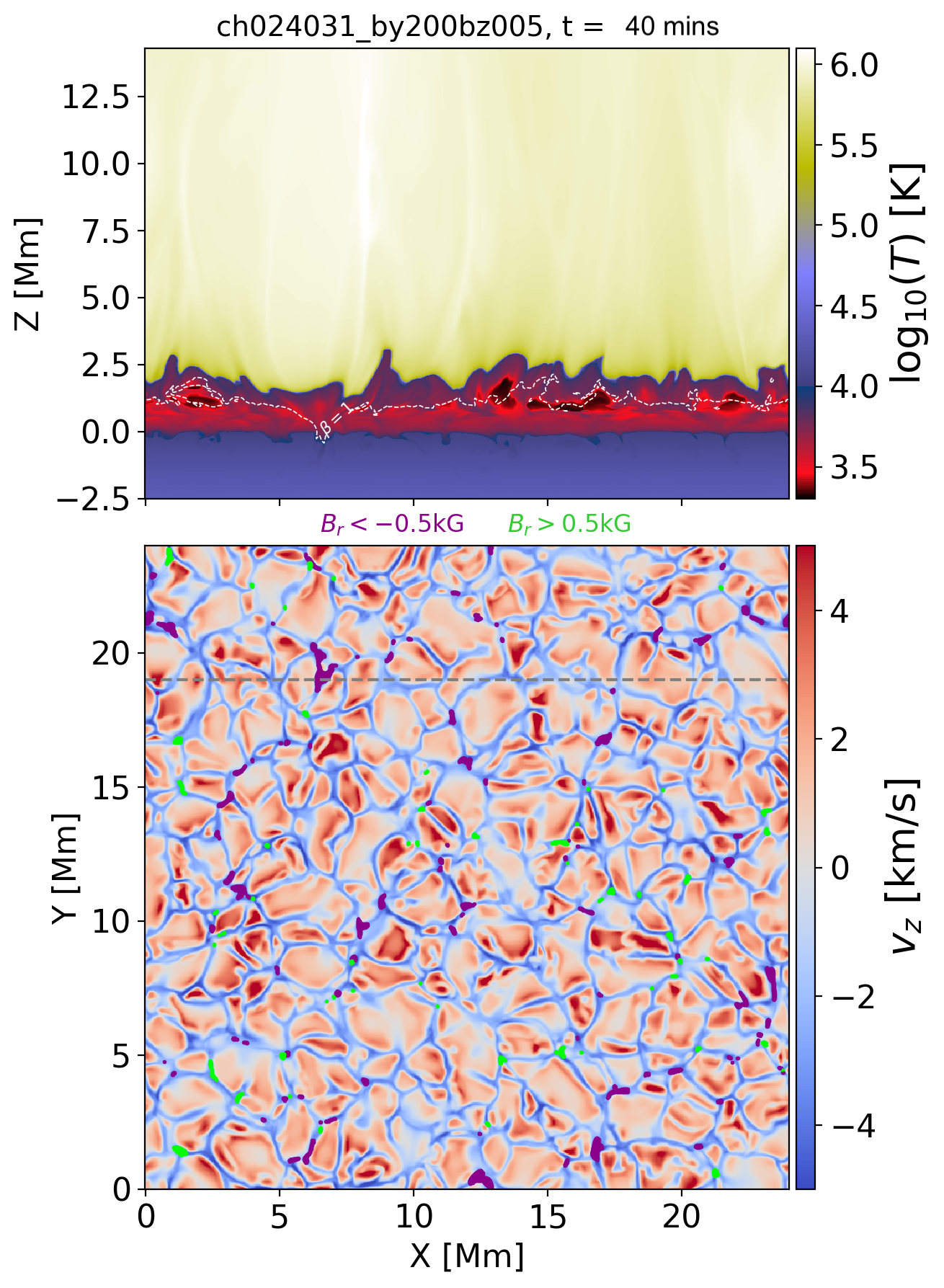}
   \caption{Top: Temperature structure in a vertical cut through the computational domain at Y = 19 Mm (indicated with a dashed line in the lower panel). The plasma $\beta$ equal to one surface is highlighted with a thin dashed line. Bottom: Vertical velocity at the photosphere (Z = 0 Mm). The strongest magnetic field enhancements are highlighted by polarity (purple = negative, green = positive). The field at the photosphere is mostly negative, and contained in the intergranular lanes. Data is shown from the same snapshot as Figure \ref{figureTempConvStructure}.}
   \label{figureTempConvStructure}
\end{figure}

\subsection{The Bifrost Code}
The simulation analysed in this study was performed with the 3D radiation magnetohydrodynamics (RMHD) code \bifrost{} \citep[described fully within][]{gudiksen2011stellar}. \bifrost{} solves the MHD equations on a staggered Cartesian grid and includes (in this case) the influence of; thermal conduction along magnetic field lines \citep{spitzer1956physics}, ohmic heating, viscous dissipation, and a range of radiative heating/cooling processes. The \bifrost{} code utilises a split-diffusive operator such that diffusive terms are separated into a small globally applicable term and a local term (often referred to as ``hyper diffusion''). This allows the code to remain highly stable and compute parameter regimes which would not be possible using a single value. This means, however, that the values of magnetic diffusivity and viscosity vary in time and space during the computation. 

To account for the radiative processes, \bifrost{} solves for the radiation field within the domain in parallel with solving the MHD equations. The radiative transfer is performed in two manners, 1) optically thin radiative transfer for the coronal/chromospheric plasma with $T>2\times 10^4$ K, which utilises a pre-computed transfer function $f(T)$, derived from tabulated atomic data in CHIANTI \citep{dere1997chianti, landi2006chianti}. This is supplemented with empirical fits from \cite{carlsson2012approximations}, which describe; the losses form the chromosphere due to strong lines, the heating from the Ly$\alpha$ line of Hydrogen, and the heating of EUV photons (thin losses from the transition region and corona). Furthermore, 2) full radiative transfer for the optically thick regions (typically in the photosphere and low chromosphere). The second method is computationally costly and so some approximations are made, i.e. a static medium, and Local Thermodynamic Equilibrium (LTE) for the gas opacity (thus the opacity $\sigma(\rho,T)$ can be pre-computed). Scattering is also included \citep{skartlien2000multigroup}, though this means that the radiative transfer equation must be solved iteratively to produce a consistent radiation field. To simplify the number of spectral lines, \bifrost{} implements an approximate opacity spectrum by replacing monochromatic opacities with mean opacities and solving the wavelength-integrated radiative transfer in four opacity bins instead \citep{nordlund1982numerical, hayek2010radiative}. 

To close the MHD equations an Equation of State (EoS) is needed. The code has a variety of EoS implementations. For this study the EoS includes 
ionization and excitation calculated in LTE from 16 elements (H, He, C, N, O, Ne, Na, Mg, Al, Si, S, K, Ca, Cr, Fe and Ni) with abundances from \citet{1975A&A....42..407G}. This (old) set of abundances was chosen in order to have the same EoS as in the relaxed 
simulations of solar convection by Stein \& Nordlund
\citep[e.g.,][]{2000SoPh..192...91S} used as a starting point for the simulations, see below.

\subsection{The Coronal Hole Patch}

\begin{figure}[t!]
 \centering
  \includegraphics[trim=1cm 0.3cm 0.5cm 0.3cm, width=0.4\textwidth]{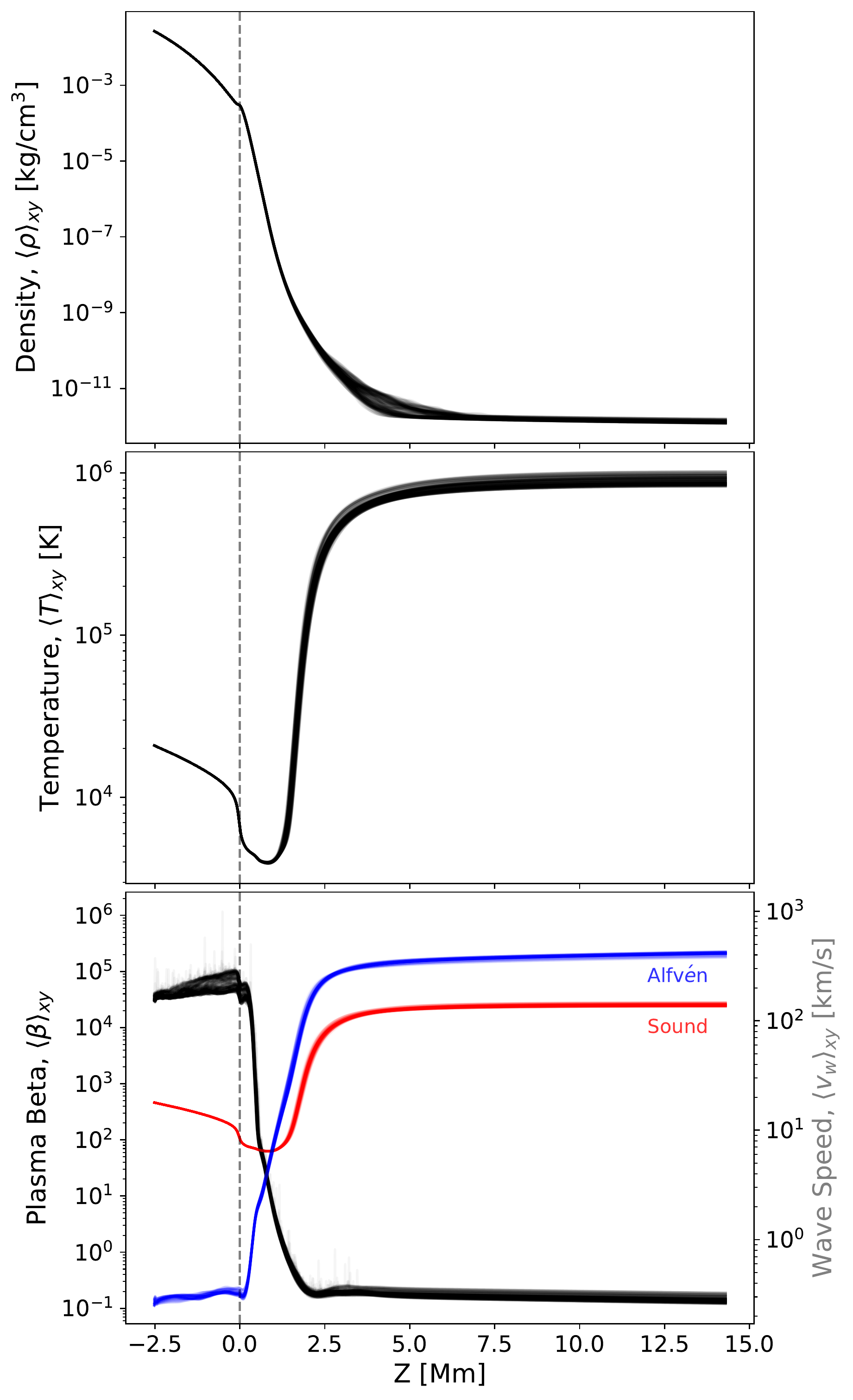}
   \caption{Average density (top), temperature (middle), plasma beta, and the sound/Alfv\'en speeds (bottom) profiles in the simulation domain for each 10 s snapshot during the hour considered. Variations in the density above the chromosphere occur due to the launching of jets. Variation in the maximum temperature results from the varying degrees of twisted magnetic structures that facilitate vertical energy transport.}
   \label{figureAverageProfiles}
\end{figure}

We examine the simulation ``\simulation{}'', which was originally produced as part of a set of simulations for the Hinode Science Data Centre\footnote{The Hinode Science Data Centre Europe (http://sdc.uio.no/search/simulations).}, intended for comparison with high resolution observations of the solar atmosphere \citep[see discussion in][]{carlsson2016publicly}. The computational domain has a horizontal extent of 24 x 24~Mm$^2$ with a resolution of dx = dy = 31~km, and spans from 2.5~Mm below the photosphere to 14.5~Mm above with a varying vertical resolution of dz = 12 - 82~km (in total 768 x 768 x 768 grid points). 

The horizontal boundary conditions are periodic, the lower and the top boundaries are open. The entropy of the 
incoming fluid at the bottom boundary is set with the aim of giving the
solar effective temperature of 5780~K. The effective temperature is not quite 5780~K. It varies in time between 5720 and 5774~K. To get it closer to 5780~K would mean changing the value of the incoming entropy and waiting for the atmosphere to relax --- a very time-consuming process.

\begin{figure*}[t!]
 \centering
  \includegraphics[trim=0.2cm 0.0cm 0.2cm 1cm,clip, width=\textwidth]{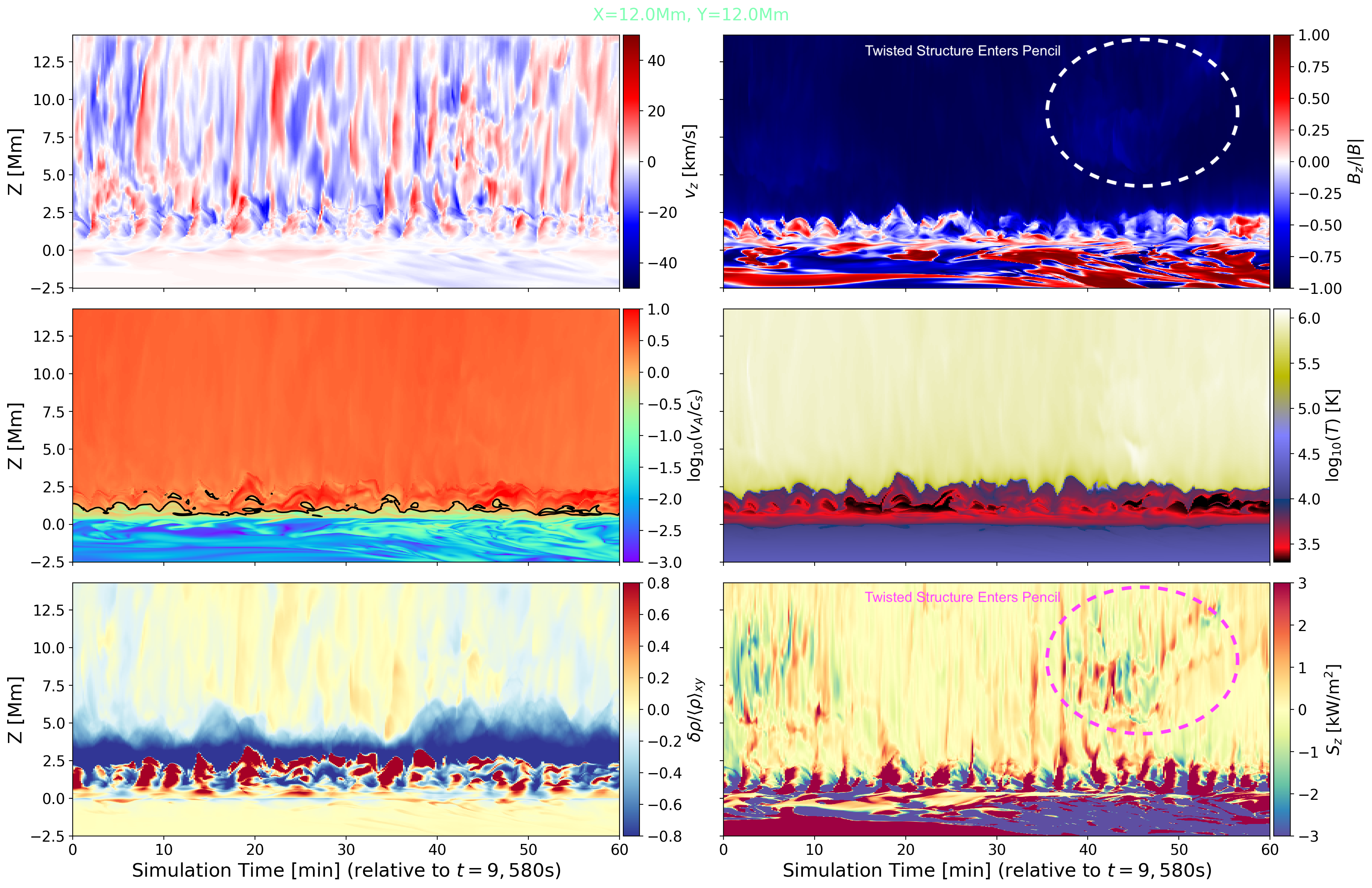}
   \caption{Time-distance plots of a vertical pencil in the simulation domain at X = 12 Mm and Y = 12 Mm. Quantities shown are; the vertical flow velocity $v_z$, the normalised vertical magnetic field $B_z/|B|$, the ratio of the Alfv\'en speed to the sound speed $v_A/c_s$, the plasma temperature $T$, the density fluctation with respect to the average of the horizontal domain $\delta\rho/\langle \rho\rangle_{xy}$, and the vertical Poynting flux $S_z$.}
   \label{figureTimeDistance}
\end{figure*}

The simulation started from a simulation cube from \citet{2000SoPh..192...91S} with a horizontal extent of 6 x 6 Mm$^2$ spanning 
from 2.5~Mm below the photosphere to 0.5~Mm above. This was duplicated
first to a 12 x 12~Mm$^2$ box and then to a 24 x 24~Mm$^2$ box and run until the
periodicities disappeared. This photospheric box (768 x 768 x 192 grid points) was run until well relaxed (14 hours of solar time). The magnetic
field in the simulation box is the result of the field in the initial
simulation cube, a small scale dynamo and the insertion of a horizontal
field with a strength of 45~G in the inflow regions. The vertical field
is balanced with zero average signed vertical field. A plane-parallel chromosphere and corona was added to this relaxed
photospheric simulation. The density and temperature structure was
taken from another simulation and the magnetic field was taken from
a potential field extrapolation from the photospheric simulation. 
This was run for 900~s after which a unipolar vertical field of
5~G was added in order to mimic the mean signed field of 
coronal holes \citep{zwaan1987elements}. The initial chromosphere and corona cools down
because there is no heating in the plane-parallel atmosphere with a potential magnetic field
to balance the optically thin radiative cooling and the energy transport to the lower atmosphere through conduction.
At the time of the addition of the vertical field, the upper atmosphere has cooled down to
about 200~kK. More and more waves and currents provide heating and the corona slowly heats up
to reach a quasi-steady temperature structure after about 8,000~s with a coronal temperature 
of 0.8--1~MK.

Figure \ref{figureFieldStructure} shows a 3D visualisation of the magnetic field in the simulation domain. The simulation contains no large-scale bipoles in the magnetic field. The majority of the magnetic flux has been swept into the intergranular lanes (see Figure \ref{figureTempConvStructure}) with an average unsigned field of 40~G. A quasi-steady temperature structure has developed inside the simulation domain, depicted in Figure \ref{figureTempConvStructure}, with a photospheric temperature around 5,780 K, cool chromosphere, and hot corona. The publicly available dataset spans approximately two hours of solar time, with snapshots available at a cadence of 10~s. We have continued this computation for an additional hour and a half of solar time, which allows more structure to develop in the computational domain via the convective braiding of magnetic field lines. A comparison of the magnetic field and temperature structure with the publicly available dataset is available in Appendix \ref{public}. For this study, we will consider the final hour of the simulation ($t=9,580-13,180$~s) and utilise a snapshot cadence of 10~s.

\section{Simulation Results}
\subsection{Overview}

\bifrost{} simulations are rich in dynamical phenomena, with this case being no different. During the hour of data selected for analysis, the magnetic field continually evolves due to the presence of convective motions at the photosphere. This triggers sporadic mass transfer into the corona, as well as braiding of the coronal magnetic field, and the modification of MHD wave propagation through the simulation domain. Despite the large number of time-varying processes, the average density and temperature profiles versus height remain roughly constant; shown in Figure \ref{figureAverageProfiles}. For this study, our interest is focused on the dynamics of the simulation and so we will not discuss in detail the various physical processes that heat and cool the plasma during the computation. 

In Figure \ref{figureTimeDistance}, a range of quantities taken from a vertical pencil at X = 12 Mm and Y = 12 Mm (i.e., the centre of the box) are plotted during the entire hour of data. Note that structures can move in and out of the selected column as the simulation is 3D. Figure \ref{figureTimeDistance} illustrates the vertical structuring of the domain, and the degree to which this evolves with time. The vertical movement $v_z$ of plasma in the corona shows plasma rising and falling at speeds of around 30 km/s, in contrast to speeds of a few km/s in the upper-convection zone. Large up-flows can supply significant amounts of material into the chromosphere (and to a lesser-degree the low-corona), as indicated by the increased density contrast $\delta\rho/\langle \rho\rangle_{xy}$ (where $\langle \rho\rangle_{xy}$ is the horizontally-averaged density $\rho$). As the field in the low-corona is unipolar and negative, the vertical extent of the closed magnetic field can be inspected from the ratio of the vertical magnetic field strength $B_z$ and the magnitude of the magnetic field vector $|B|$; i.e. $B_z/|B|$ (indicating the degree to which the field is vertical). The closed field is generally contained below Z = 2.5 Mm (identifiable by $B_z/|B|\geq 0$). This coincides with the top of the chromosphere, which is visible in the plasma temperature $T$. 

The sound and Alfv\'en wave speeds are a crucial diagnostic of the state of the atmosphere, these are given by,
\begin{equation}
    c_s = \sqrt{\frac{\partial P}{\partial\rho}} = \sqrt{\frac{\gamma P}{\rho}},
\end{equation}
and
\begin{equation}
    v_A = \frac{B}{\sqrt{\mu_0 \rho}},
\end{equation}
respectively, where $P$ is the gas pressure, $\gamma$ is the adiabatic index, and $\mu_0$ is the vacuum permeability. The average profiles of these wave speeds versus altitude in the simulation domain are shown in Figure \ref{figureAverageProfiles}. The surface where the ratio of the Alfv\'en speed to the sound speed $v_A/c_s$ is equal to one, is approximately the plasma beta $\beta$ equals one surface and is highlighted with a black line in the time-distance plot of $v_A/c_s$ ($\sim 1/\sqrt{\beta}$) in Figure \ref{figureTimeDistance}. The plasma beta is given by,
\begin{equation}
    \beta = \frac{P}{B^2/2\mu_0} =\frac{2}{\gamma} \frac{c_s^2}{v_A^2}.
\end{equation}
Below the $\beta=1$ surface, the plasma pressure dominates the magnetic pressure, and vice versa above. This surface is located around 1 - 1.5 Mm above the photosphere. Thus the convection efficiently influences the magnetic field at the photosphere, whereas the magnetic field (relaxation) dominates the dynamics in the low-corona (discussed in Section \ref{braiding}). As well as indicating the changing pressure regime, the $v_A/c_s=1$ surface is also a location of significant interest for mode conversion (discussed in Section \ref{waves}). 

\begin{figure*}[t!]
 \centering
  \includegraphics[trim=0.0cm 0.0cm 0.0cm 0cm,clip, width=\textwidth]{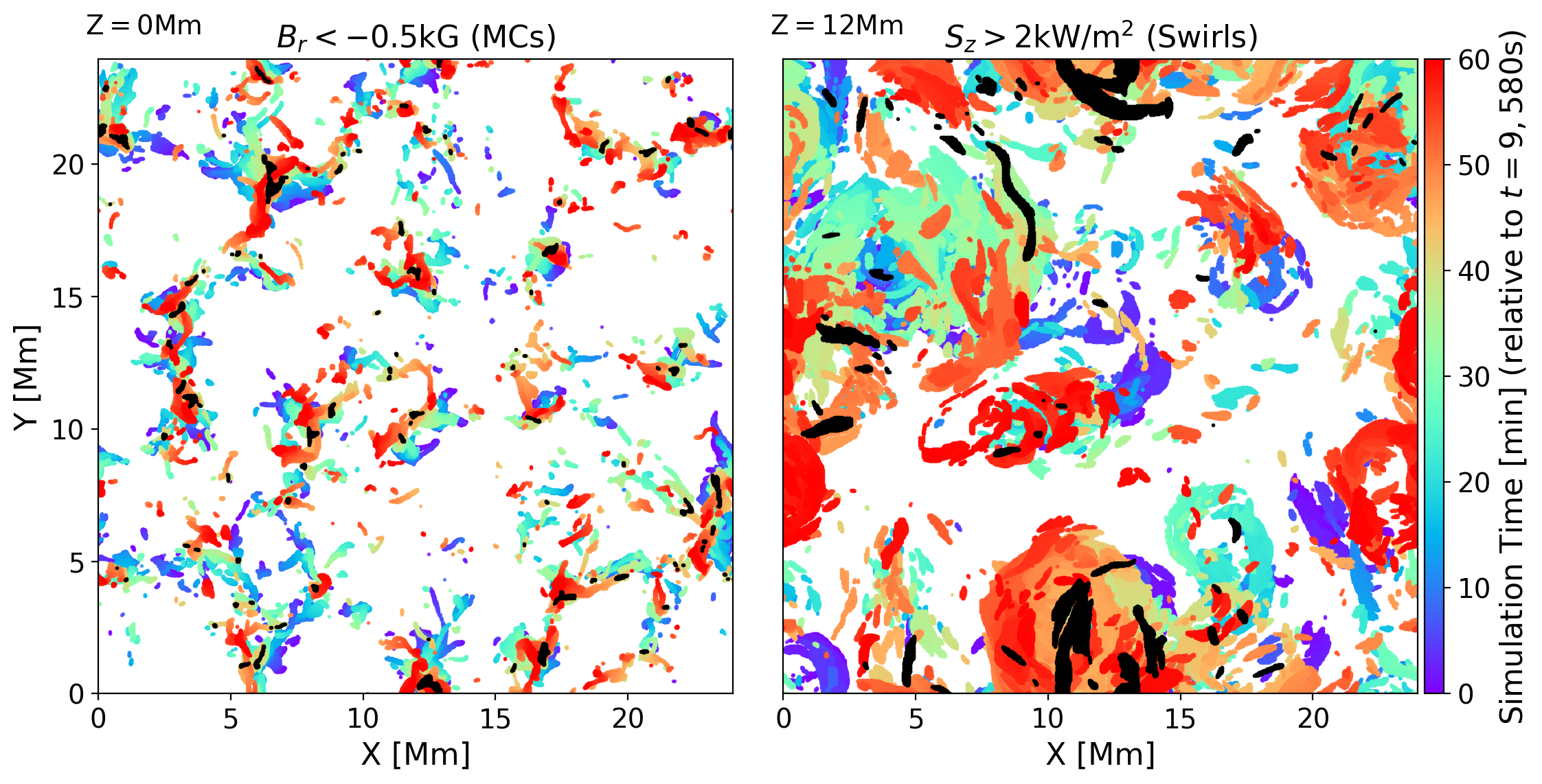}
   \caption{Left: The location of Magnetic Concentrations (MCs) at the photosphere (Z = 0 Mm) during the hour of simulation time. Right: The location of strong vertical Poynting fluxes in the low-corona (Z = 12 Mm), associated with the braiding and twisting of the magnetic field by convective motions, and the release of energy in Swirling Events (SEs). The spatial distribution of the MCs and Swirls at $t=40$~mins are highlighted in black.}
   \label{figureMEsAndSwirls}
\end{figure*}

\begin{figure}[t!]
 \centering
    \includegraphics[trim=0.2cm 0.3cm 0.3cm 0.2cm,clip, width=0.48\textwidth]{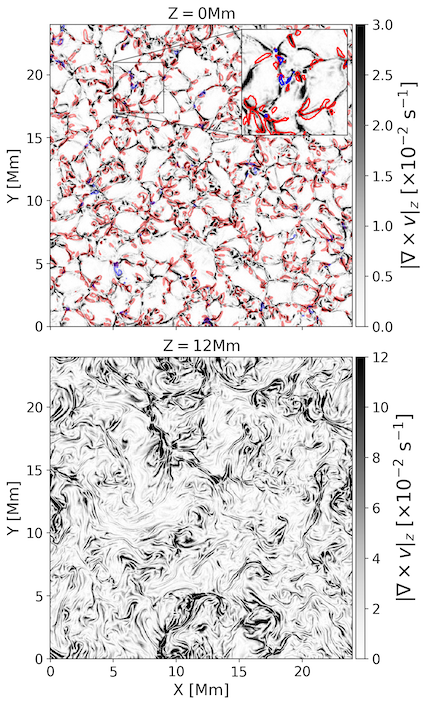} 
   \caption{Horizontal cuts of the vorticity magnitude $|\nabla\times{\bf v}|_z$ at Z = 0 Mm and Z = 12 Mm, for $t=40$~mins. Contours of the kinetic helicity ${\bf v}\cdot(\nabla\times{\bf v})$ (red) and current helicity ${\bf B}/(\mu_0\rho)\cdot(\nabla\times{\bf B})$ (blue) at a value of 100 m/s$^2$ are plotted with the vorticity in the Z = 0 Mm panel. The zoomed inset within the first column (Z = 0 Mm) is 4 Mm x 4 Mm, centered on X = 7 Mm and Y = 19 Mm. The same patch of the domain is discussed again in Figure \ref{figureSwirling}.}
   \label{figureVorticity}
\end{figure}

\begin{figure*}[t!]
 \centering
  \includegraphics[trim=0.0cm 0.0cm 0.0cm 0cm,clip, width=\textwidth]{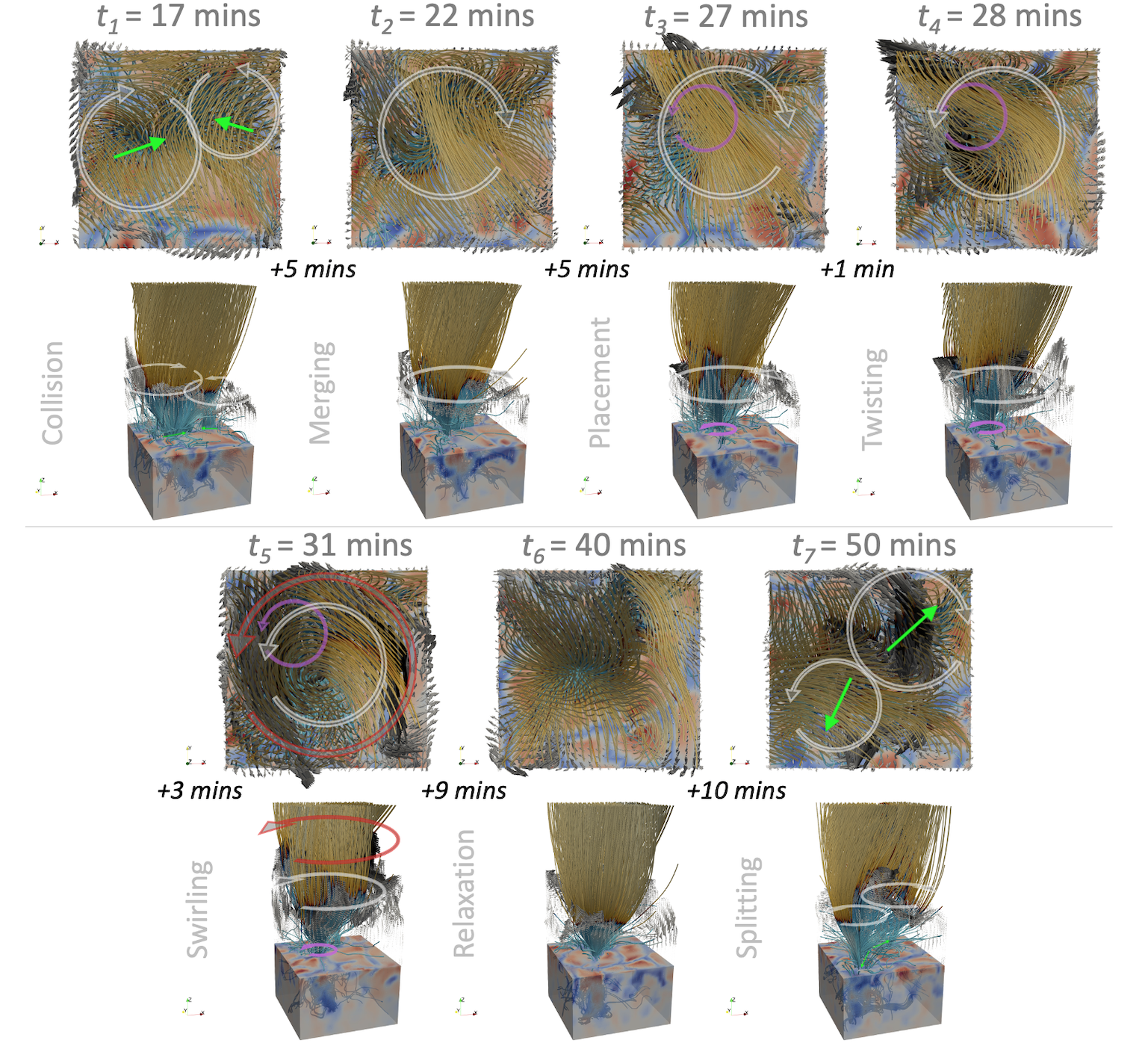}
   \caption{Snapshots during a $\sim$30 minute period that display the merging of two MCs within the intersection of multiple granules, and the subsequent excitation of a SE. A subsection of the computational domain is shown that extends from the bottom boundary up to 6 Mm above the photosphere, with a horizontal extent of 4 Mm x 4 Mm (centered on X = 7 Mm and Y = 19 Mm). A top-down view of the domain is shown above each snapshot. The same cut in the domain is highlighted in Figure \ref{figureVorticity}. The upper-convection zone is coloured by vertical velocity (red is rising, blue is falling material). Selected magnetic field lines are coloured by temperature as in Figure \ref{figureFieldStructure}. The flow velocity in the chromosphere is indicated with arrows, the size and colour of which depend on the magnitude of the local flow speed. Cartoon annotations indicate the direction of advection for the MCs (green) and} the important/prevailing flow patterns in the photosphere (magenta), chromosphere (white), and low-corona (red).
   \label{figureSwirling}
\end{figure*}

\begin{figure*}[t!]
 \centering
  \includegraphics[trim=0.4cm 0.3cm 0.5cm 0.3cm, width=\textwidth]{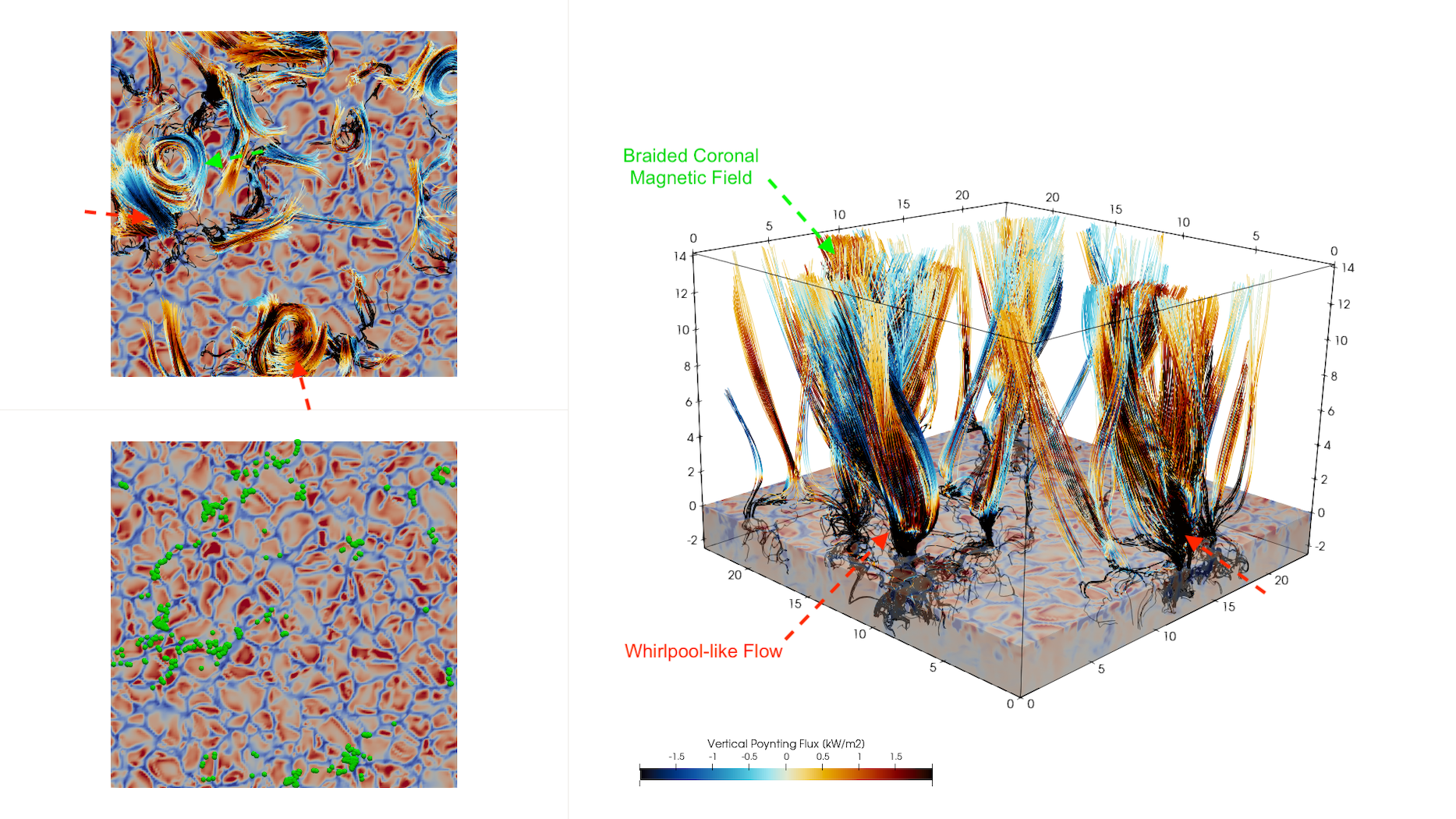}
   \caption{Right: 3D rendering of the twisted magnetic field structures in the simulation domain at $t=40$~mins, identified with large values of Poynting flux ($|S_z|>2$ kWm$^{-2}$) at Z = 6 Mm. Top left: As viewed from above. Bottom left: The location of the magnetic field footpoints at the photosphere (Z = 0 Mm) are highlighted with green markers.}
   \label{figureTwistedFluxTubes}
\end{figure*}

\subsection{Twisting and Braiding the Coronal Magnetic Field}\label{braiding}
There are many routes to forming twisted structures in the solar atmosphere. It is known that convective motions can generate vorticity in-between neighbouring convective granules \citep{giagkiozis2018vortex}, and that this can twist-up the magnetic field rooted in the intergranular lanes (when $\beta$>>1). Similarly, \cite{fischer2020interaction} found evidence for the generation of twisted horizontal structures, by shallow recirculation near the photosphere. In addition to braiding at the solar surface, the emerging flux may already have become twisted as it rose through the upper-convective zone \citep{pinto2013flux, mactaggart2021helicity, mactaggart2021direct}. In the \simulation{} simulation, the braiding of field lines and generation of structure in the low-corona is principally driven by the convective motions dragging the footpoints of the magnetic field around the photosphere. The coronal magnetic field in the simulation is structured into many distinct magnetic funnels. These funnels originate from Magnetic Concentrations (hereafter MCs) in the photosphere that expand with height into the low-corona where they jostle with other funnels to fill the available volume. This configuration has been well documented \citep{hofmeister2019photospheric} and discussed in the literature \citep[see][and references therein]{tu2005solar}. 

These funnel networks are thought to evolve slowly, with theoretical studies often examining the dynamics of simplified (often 2D) funnel configurations \citep{pascoe2014fast, wojcik2019two}. In the \bifrost{} simulation, the MCs (which are the sources of the funnels in the photosphere) move around the domain relatively slowly (up to $\sim5$ Mm/hour), matching the average horizontal flow velocity at the photosphere of $\sim 2$ km/s. For reference, the average Alfv\'en and sound speeds at the photosphere are $\sim0.5$ km/s and $\sim8$ km/s, respectively. The MCs migrate around the intergranular lane network, where the magnetic flux of each MC can be broken apart and shuffled around. There are around 50 distinct MCs at any time in the simulation, that contain an average unsigned magnetic flux of $1.5\times 10^{16}$ Mx, and area of $4\times 10^{-2}$ Mm$^2$ (a cross-section of $\sim 0.11$ Mm). These MCs are associated with an average kinetic helicity ${\bf v}\cdot(\nabla\times{\bf v})$ of $50$ m/s$^2$ (with current helicity ${\bf B}/(\mu_0\rho)\cdot(\nabla\times{\bf B})$ of $4$ m/s$^2$). The shuffling of MCs slowly adds complexity to the coronal magnetic field by braiding together different magnetic field lines from various locations across the photosphere. Figure \ref{figureMEsAndSwirls} shows the locations of MCs in the photosphere, coloured by simulation time (beginning blue and finishing red), which shows their confinement in the intergranular lanes. During extreme events, localised vortex-flows in the photosphere can rapidly twist the magnetic field, on timescales of 30~s to a few minutes \citep[seen also in observations][]{bonet2010sunrise}. 

Figure \ref{figureVorticity} shows the magnitude of vorticity at the photosphere (Z = 0 Mm) and the low-corona (Z = 12M m). At the photosphere, hot material rises up inside the granular convection after which it cools and is displaced towards the intergranular lanes. These flow merge in the intergranular lanes where the density becomes enhanced and vortical flows are generated \citep[an initial source of vorticity is required to generate a true whirlpool down-flow, see][]{simon1997kinematic}. Large values of the kinetic helicity are highlighted with red contours, and follow the structure of the intergranular lanes. The current helicity is also enhanced at the base of the magnetic funnel structures, where the field strengths are largest. Areas with high current helicity, blue contours in Figure \ref{figureVorticity}, are more susceptible to forming the strong whirlpool-like down-flows. The flow vorticity is largest in the chromosphere above, as the density decreases and so the flow speeds can be consequently larger for the same amount of energy. The MCs, which expand with height to form the magnetic funnel network, are regularly buffeted by the convective/vortical motions. This launches torsional Alfv\'en waves up and along the funnels, which results in the vorticity of the low-corona being highly structured and complex (see Figure \ref{figureVorticity}). The propagation of these fluctuations will be discussed in Section \ref{waves}.

Twisting motions in the magnetic field, driven by the photospheric convection, transfer magnetic energy into the low-corona as a Poynting flux \citep[see discussion of][]{hansteen2015numerical}. The vertical Poynting flux is given by,
\begin{equation}
    S_{z}=\frac{1}{\mu_0}\bigg({\bf E}\times{\bf B}\bigg)_z,
\end{equation}
where $\bf E$ is the electric field given by ${\bf E} = - {\bf v}\times {\bf B} + \eta{\bf J}$. The current density is given by ${\bf J} = 1/\mu_0 \nabla\times{\bf B}$. The vertical Poynting flux $S_z$ can be broken down into three terms, the first being the emerging Poynting flux,
\begin{equation}
    S_{z,fe}= \frac{v_z}{\mu_0}(B_x^2+B_y^2),
\end{equation}
the second being the field shaking/shearing Poynting flux,
\begin{equation}
    S_{z,fs}= \frac{-B_z}{\mu_0}(v_xB_x+v_yB_y).
\end{equation}
and the third resulting from the the finite magnetic diffusivity $\eta$,
\begin{equation}
    S_{z,\eta}= \frac{\eta}{\mu_0^2}\bigg[\bigg(\frac{\partial B_z}{\partial y}-\frac{\partial B_y}{\partial z}\bigg)B_x+\bigg(\frac{\partial B_x}{\partial z}-\frac{\partial B_z}{\partial x}\bigg)B_y\bigg].
\end{equation}
The sum of these terms equals the total vertical Poynting flux, i.e., $S_z$ = $S_{z,fe}$ + $S_{z,fs}$ + $S_{z,\eta}$. The term $S_{z,fe}$ is associated with the advection of horizontal magnetic field by a vertical flow, and the term $S_{z,fs}$ is linked with the horizontal motion (shaking/shearing) of vertically aligned field (this is how energy is transported by Alfv\'en waves, further discussed in Section \ref{alfvenwaves}). The term $S_{z,\eta}$ relates to the dissipation of currents and is negligible ($|S_{z,\eta}|<<10^{-10}$ W/m$^2$) such that $S_z$ $\approx$ $S_{z,fe}$ + $S_{z,fs}$.  In general, the emergence term dominates the average Poynting flux in the convective zone (as vertical fluid motions control the motion of the field) and is typically negative due to the subduction of magnetic field in intergranular lanes. Above the photosphere the shaking/shearing term dominates and is on average positive (though its strength diminishes with altitude). The location of Poynting flux enhancements at Z = 12 Mm are shown in Figure \ref{figureMEsAndSwirls}, in comparison to the driving MCs in the photosphere. Upon visual inspection, the locations of Poynting flux enhancements and the movements of MCs across the photosphere are related. However the low-corona magnetic field is deformed from purely vertical, due to its expansion, so magnetic energy is not always deposited directly above the driving MCs. The largest and most complex network of MCs, with braided coronal field, almost continuously heat the low-corona above them during the hour of simulation time. This has interesting implications for the solar wind above, that are discussed in Section \ref{structure}.

For the majority of interactions between the convection and the photospheric magnetic field, a moderate twist is applied to the field and this magnetic energy can be transmitted into the low-corona by a steady Poynting flux. The unravelling of these twisted structures is often resisted/delayed by the accumulation of chromospheric plasma that feels a reduced ``effective'' gravitational force due to the increased horizontal component of the magnetic field. Thus the chromosphere can sometimes thicken inside these twisted structures. However, when the convective motions drive MCs into complex intersections between multiple granules, they can remain in place long enough to influence the magnetoconvection. Often this creates a whirlpool-like down-flow, as material sinks towards the center of the depression whilst conserving angular momentum \citep{kitiashvili2012dynamics, moll2012vortices, silva2020solar}. Whirlpool-like photospheric flows twist the magnetic field too quickly for it to remain in equilibrium with its surroundings. This causes the field to dramatically rotate/swirl which we refer to as a \textit{swirling event} (SE).

Figure \ref{figureSwirling} follows the merging of MCs at an intersection of granulation, and the subsequent generation of a SE in the low-corona, during a 30 minute interval. At $t_1$, two neighbouring flux bundles, that are embedded along the same intergranular lane, are forced to merge due to random convective motions. The movement of the magnetic funnels generates a vortical flow around the flux bundle in the chromosphere. At $t_2$, the flows around each funnel merge and cancel out some of the vorticity associated with one of the flux bundles. The remaining flow encircles the recently merged flux, allowing the bundles to merge. During $t_3$, the newly formed MC is compacted by the converging granulation flows and placed in the intersection of granulation, which produces a localised depression in the photosphere. A whirlpool-like flow begins to form in the depression.  At $t_4$, the field starts to be twisted by the photospheric flow. The coronal field begins to slowly twist (transferring a steady Poynting flux into the low-corona) which (in this case) reverses the pre-existing circulation of the chromospheric flow to match that of the photosphere. The magnetic field generates a horizontal component due to the twisting which can support additional chromospheric plasma, and so the chromosphere thickens with the additional material resisting the twisting of the field (storing magnetic energy). At $t_5$, the whirlpool-like flow in the photosphere has strengthened and rapidly twists the MC. The field above is stressed too quickly to remain stable (typically the low-corona magnetic field rotates violently one to two minutes after the whirlpool formation). The magnetic tension force dominates in the chromosphere, and the entire magnetic funnel structure begins to rotate, launching an Alfv\'enic pulse from the release of the stored magnetic energy. This disturbance propagates up along the magnetic field lines (ejecting chromospheric plasma into the low-corona) followed by a swirling motion, driven directly by the convection. At $t_6$ the MC leaves the intersection, and becomes less concentrated. The overlying field can now relax, returning to a single magnetic funnel anchored to a MC. As the chromospheric plasma has been largely ejected, the plasma pressure within the funnel is low, in addition to the $\beta=1$ surface being depressed due to the suppression of magnetoconvection (see Figure \ref{figureTempConvStructure}). Finally, at $t_7$ the magnetic funnel is slowly split apart by the convective motions, breaking apart the MC. Additional energetic events are driven by the merging and splitting of the magnetic funnels, though here only the dominant swirling event has been discussed. Despite the event lasting around 30 minutes from start to finish, the SE (the release of built-up magnetic energy) lasted around 10 minutes, beginning shortly after $t_4$ and reaching a relaxed state at $t_6$. 

\begin{figure}[t!]
 \centering
  \includegraphics[trim=0cm 0cm 0cm 0cm, width=0.45\textwidth]{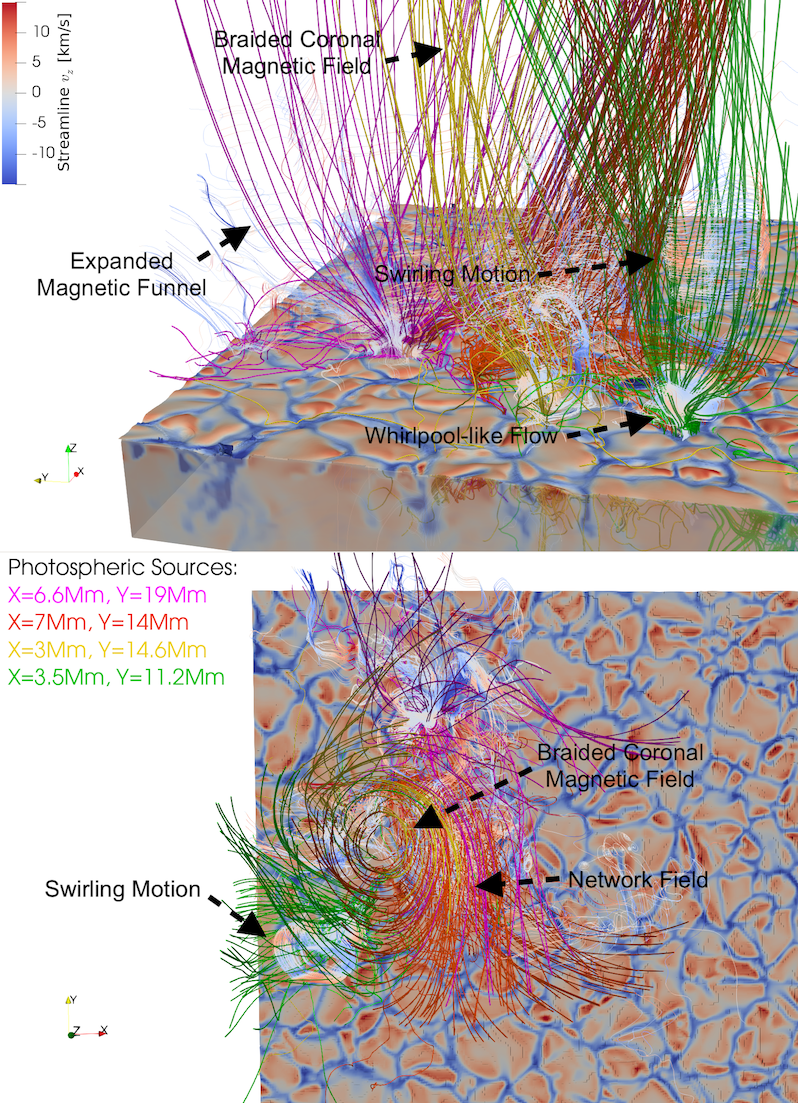}
   \caption{Top: A 3D visualisation of a subset of the braided coronal magnetic field, at $t=40$~mins. Magnetic field lines are traced from four photospheric sources (seed points are initiated within 0.25 Mm of the given coordinates). Field lines are coloured by their source location. Streamlines of the velocity field are initiated at the photospheric sources, and are coloured by vertical flow velocity $v_z$. Bottom: View from above. }
   \label{figureColouredSwirls}
\end{figure}

As twisted/braided magnetic field lines carry significant Poynting fluxes into the low-corona, this is used as a criterion to highlight these structures inside the computational domain. Figure \ref{figureTwistedFluxTubes} shows a snapshot of all the magnetic field lines which have $|S_z|>2$ kWm$^{-2}$ at a height of 6 Mm. This criterion is chosen as it highlights the majority of twisted structures in the computational domain. The photospheric footpoints of the field lines identified by this criterion are indicated with green markers. From this visualisation, it is clear that there are a range of stressed-structures in the domain, with a variety of sizes and shapes. Each magnetic funnel that has been twisted or braided together has varying degrees of stressed field lines inside, thus the structure in the Poynting flux is often of a finer-scale than the overall funnel-size. The fine-structure in the Poynting flux is observed in Figure \ref{figureTimeDistance} from $t=35$~mins to $t=55$~mins (circled with a dashed magenta line), when a twisted funnel structure enters the vertical cut (the signature of this is the ratio of $B_z/|B_z|$ decreasing). In Figure \ref{figureTwistedFluxTubes}, the Poynting flux enhancements are not always positive (both blue and yellow tones are visible). This is largely due to two factors. Firstly the sinking of coronal plasma back to the photosphere, and secondly the unwinding of the magnetic field in the direction of the photosphere (often observed as a collapse in the center of large structures). A clear trend emerges when viewing a timeseries of visualisations like Figure \ref{figureTwistedFluxTubes}, the more magnetic field is braided together from different MCs (i.e., increasing the complexity of the structure), the more time it will take to dissipate/unwind. The smallest stressed bundles of field lines, which do not have a particular shape, come and go from the detection criterion of $|S_z|>2$ kWm$^{-2}$, in around 20~s to a few minutes. Whereas small vortex-like structures above isolated MCs (2 - 5 Mm in diameter in the low-corona) can remain visible for around 5 - 10 minutes. The most complex structures ($\sim$ 10Mm in diameter), resembling multiple structures braided together, can remain for up to 30 - 40 minutes at a time. 

The field lines highlighted by this criterion have photospheric footpoints in the intergranular lanes with MCs, as expected. The largest structures, are composed of tangled magnetic field from a variety of photospheric sources. To show this, Figure \ref{figureColouredSwirls} displays another visualisation of the same snapshot, now focusing on the largest twisted structure in the domain. Field lines from four photospheric sources are drawn, each with a different colour.  Due to the shuffling of MCs around the intergranular lanes, the magnetic field from the four sources is tangled together into a large-scale braided field. Field lines in magenta depict a relaxed magnetic funnel, this is the typical configuration of the magnetic field emerging from MCs. The green field lines highlight a magnetic funnel that has whirlpool-like motions currently at its base, and so is being actively stressed (see Figure \ref{figureTwistedFluxTubes}, where this structure is also annotated). Streamlines inside the green funnel show the plasma inside the funnel is swirling around, and that material is sinking into the whirlpool at the photosphere. The magnetic field highlighted in red corresponds to a patch of network field, i.e., field not originating exactly from the MCs and instead having a more diffuse origin (centered on the given coordinates). It is thought that the presence of this network field may be responsible for the formation of the large-scale braiding here. This field acts like an anchor, allowing the other elements to be wrapped around it by the shuffling of MCs in the intergranular lanes. 

\begin{figure*}[t!]
 \centering
    \includegraphics[trim=0.0cm 0.3cm 0.3cm 0.0cm,clip, width=\textwidth]{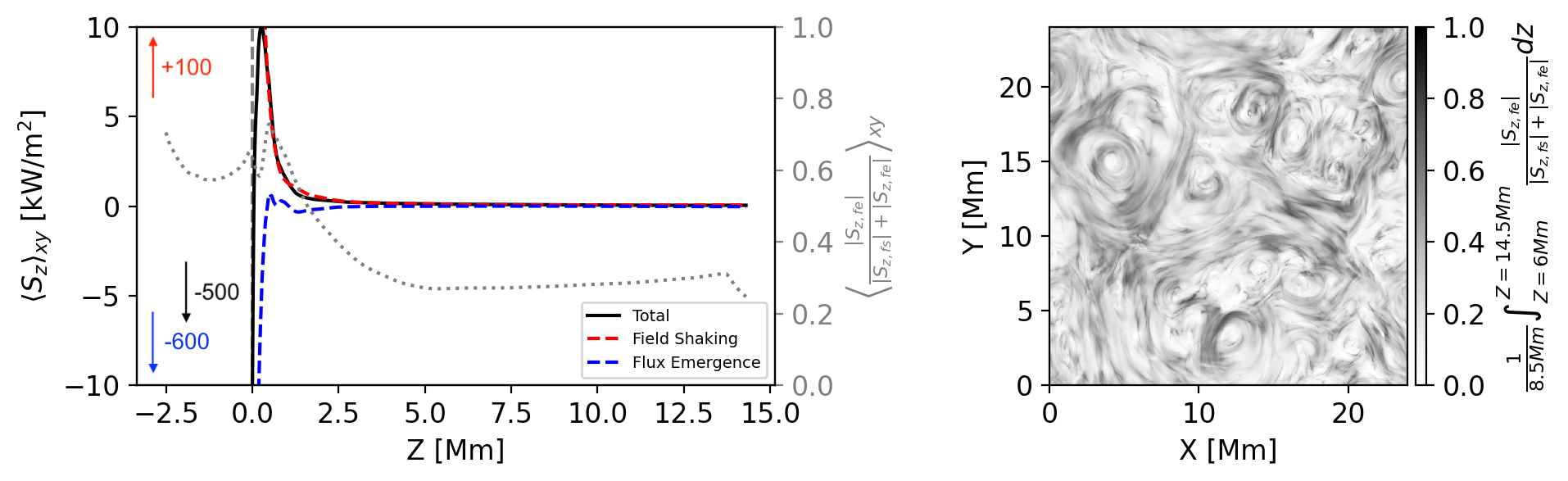} 
   \caption{Left: Average total Poynting flux versus height in black, with the field shaking $S_{z,fs}$ and flux emergence $S_{z,fe}$ terms in red and blue respectively, at $t=40$~mins. Due to the range of scale, the y-axis is limited to a few kW/m$^2$. The floating values indicate the approximate value of each term in the upper-convection zone. The average contribution of the flux emergence term to the vertical Poynting flux versus height is also shown. Right: This ratio is instead vertically-averaged above 6 Mm (i.e. at the top of the simulation domain), again for $t=40$~mins. This highlights the network of magnetic swirls/funnels that exist throughout the domain. }
   \label{figureDissipation}
\end{figure*}

Given the orientation of the magnetic field, the twisting and un-twisting of funnel structures (and the larger braided structures) appears as a flux emergence in the low-corona. So the stirring of the low-corona by convective motions is examined further by evaluating the significance of the emerging Poynting flux term to shaking/shearing term in the Poynting flux. The horizontally-averaged values of the Poynting flux terms are shown in Figure \ref{figureDissipation} for a single snapshot, along with the ratio $|S_{z,fe}|/(|S_{z,fe}|+|S_{z,fs}|)$, which weighs the relative strength of the two terms. The vertically-averaged value of this quantity in the low-corona (above 6 Mm) plotted to the right. As previously discussed, the shaking/shearing term dominates the quasi-steady background of fluctuations and so the average at the top of the simulation domain is largely grey/white. However the unwinding of the magnetic field is highlighted with darker tones (including its fine-structure). Using this criterion, the stirring of the low-corona is far more continuous than indicated by examining the extreme values of $S_z$. These un-winding motions are prevalent throughout the entire low-corona. The overall pattern changes very slowly during the hour under investigation, as this relates to the buffeting of the magnetic funnel network by convective motions at its base. Thus the pattern evolves on the timescale of the convection, and only when MCs are able to generate whirlpool-like flows in the photosphere, are the strongest Poynting flux enhancements triggered (the darkest, time-dependent features). In the low-corona, the dissipation of these stirring motions by Ohmic and viscous dissipation is low. Dissipation is largest at the boundaries of the twisted magnetic funnels, however these values are orders of magnitude smaller than the Poynting flux contained inside the twisted magnetic fields.

The magnetic energy released from a SE is primarily converted into the kinetic energy of the ejected chromospheric plasma, and the heating of that plasma to coronal temperatures. The kinetic energy density of chromospheric plasma inside a funnel structure with an active SE is enhanced by a factor of $\sim 5$ compared to the background chromospheric plasma. In the low-corona, the average plasma temperature inside a magnetic funnel connected to a SE is $\sim$0.2 MK larger than the average background temperature (of $0.9-1.0$ MK). The twisted structures in the chromosphere and the low-corona have typical diameters of 2 - 5 Mm and 4 - 11 Mm, respectively (as the funnel expands into the low-corona). However, the enhancements in $S_z$ are generally limited to an area of $\sim$2Mm$^2$ inside these structures. 

It is difficult to truly distinguish between the SEs and the more gradual twisting of the field, as they form a continuous spectrum of events with a variety of energies and sizes. The physical size of an individual SE is connected to the amount of flux inside the MC at its base, and the expansion of the magnetic funnel above. The rate at which they occur is related to how frequently the random convective motions force MCs into the intersections of granulation, where whirlpool-like flows are more readily formed. If for example, the background magnetic field in the simulation domain were stronger, we may expect the movement of the MCs through the intergranular lanes to be slower (as the dynamics would be dominated more by the magnetic field), and thus find a reduced global braiding to the coronal magnetic field. However, slightly stronger magnetic fields may increase the likely-hood of generating depressions in the convection around MCs, and subsequently produce more whirpool-like flows which locally twist the coronal magnetic field. Clearly, the properties of the structures formed in the simulation domain, from the global braiding to the SEs, are linked to magnetic field structure in the low-corona, and the convection set-up, and so care must be taken in their interpretation.

\subsection{Tracing Wave Injection into the Low-Corona}\label{waves}

\begin{figure*}[t!]
 \centering
  \includegraphics[trim=0.2cm 0.0cm 0.3cm 0.0cm,clip,width=\textwidth]{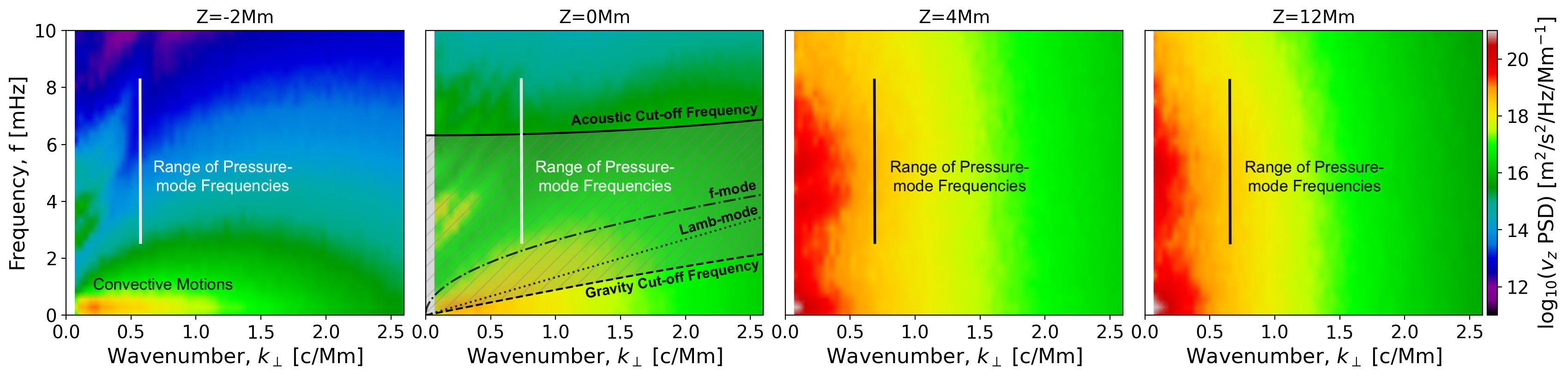}
   \caption{Power Spectral Density (PSD) in wavenumber-frequency space for the vertical flow speed $v_z$, taken at various heights in the computational domain; Z = -2 Mm (upper-convection zone), Z = 0 Mm (photosphere), Z = 4 Mm (above the chromosphere), and Z = 12 Mm (low-corona). For the photospheric slice, the typical acoustic-gravity wave cut-off frequencies are indicated, given the average pressure, density, and magnetic field strength. Classically, waves in the shaded region are evanescent (under the assumption of a stably-stratified atmosphere). For completeness the lamb-mode ($\omega = c_s k_{\perp}$) and $f$-mode ($\omega = \sqrt{g k_{\perp}}$) dispersion relations are shown with dotted and dash-dotted lines respectively. Elsewhere, the pressure-modes generated due to the computational set-up are highlighted. These ``box-modes'' have a clear impact on the PSD of the velocity fluctuations up in the low-corona.}
   \label{figureVzFFT}
\end{figure*}

\begin{figure*}[t!]
 \centering
  \includegraphics[trim=0.2cm 0.0cm 0.3cm 0.0cm,clip,width=\textwidth]{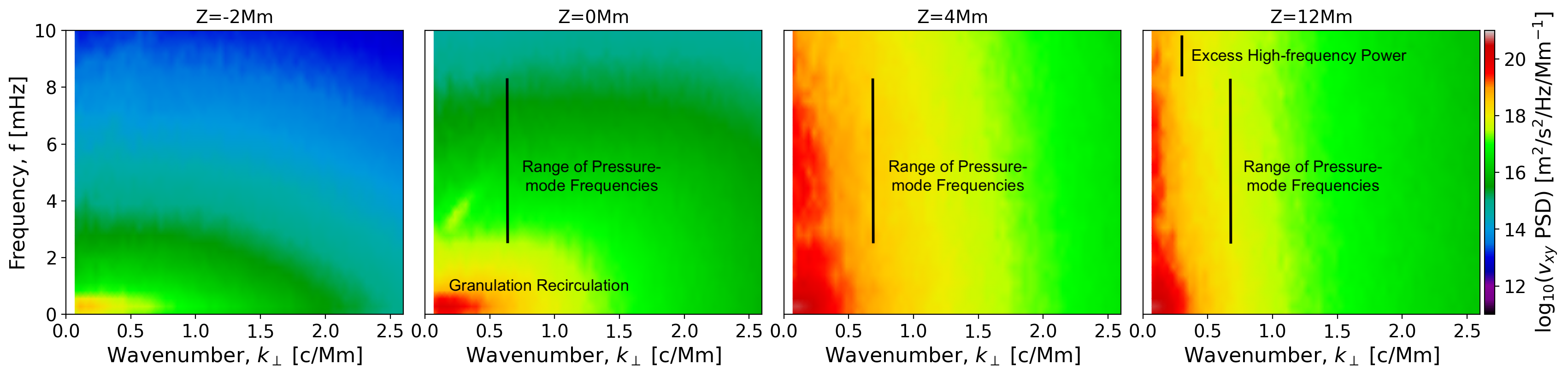}
   \caption{Same as Figure \ref{figureVzFFT}, but now for the horizontal flow speed $v_{xy}$. Enhancements in the horizontal velocity PSD at the box-mode frequencies/wavenumbers are highlighted. }
   \label{figureVxyFFT}
\end{figure*}

As the coronal magnetic field is highly twisted, and dynamically evolving, the vertical propagation of MHD waves should be affected. Previous works, that experimented with a range of photospheric drivers at the base of simplified magnetic funnels \citep{bogdan2003waves, mumford2015generation}, have shown that torisional oscillations preferentially produce Alfv\'enic wave modes (as observed during the SEs), whereas vertical perturbations excite magnetosonic waves, and horizontal driving favours kink modes. A preliminary assessment of the wave-action in the domain is made by performing Fourier Transforms\footnote{For all the frequency-based analysis in this study, we make use of the \textit{numpy} and \textit{scipy} python packages and the subroutines within.} on the spatial and temporal evolution of vertical velocity $v_z$ and horizontal velocity $v_{xy}=(v_x^2+v_y^2)^{0.5}$ in horizontal cuts through the simulation at varying heights. The components of velocity, initially described as $v_i(x,y,t)$, are transformed into their resulting Power Spectral Density (PSD) by the discrete Fourier transform,
\begin{equation}
\begin{aligned}
    & F_i(k_x,k_y,\omega) = \\ 
    & \sum_{n_x=1}^{X}\bigg(\sum_{n_y=1}^{Y}\bigg(\sum_{n_t=1}^{T} v_i(n_x,n_y,n_t) w(n_t) e^{i\frac{2\pi\omega n_t}{T}} \bigg) e^{-i\frac{2\pi k_x n_x}{X}}\bigg) e^{-i\frac{2\pi k_y n_y}{Y}},
\end{aligned}
\end{equation}
where $w(n_t)$ is a Hanning window function \citep{blackman1958measurement} of duration 30 minutes, $n_x$, $n_y$, and $n_t$ represent the indices of the spatial and temporal directions which have lengths $X$, $Y$, and $T$ respectively. Thus the wavenumbers and frequencies ($f=\omega/2\pi$) sampled are in steps of $\delta k_x =2\pi/(Xdx)$, $\delta k_y=2\pi/(Ydy)$, and $\delta \omega=2\pi/(T\delta t)$. The smallest wavenumber sampled is 0.08 cycles/Mm and the smallest frequencies are 0.3 mHz. The discrete Fourier transform $F_i$ can then be multiplied by its conjugate $F_i^*$ to recover the PSD as,
\begin{equation}
    \text{PSD}_i(k_x,k_y,\omega) = \frac{F_i(k_x,k_y,\omega) \cdot F_i^*(k_x,k_y,\omega)}{\delta k_x \delta k_y \delta\omega}.
\end{equation}
The $k_x$ and $k_y$ directions are further quadratically-summed to produce $k_{\perp}$. The PSD for the two components of velocity ($v_z$ and $v_{xy}$), at four different heights in the domain (-2 Mm, 0 Mm, 4 Mm, and 12 Mm), are depicted in Figures \ref{figureVzFFT} and \ref{figureVxyFFT}. In the upper-convection zone and at the photosphere, the high plasma $\beta$ enables the convective motions to generate acoustic waves. These waves are associated with two characteristic frequencies, the Brunt-Vaisala (or buoyancy) frequency $N$, and the acoustic cutoff frequency $\omega_c$. The Brunt-Vaisala frequency and acoustic cutoff frequency are related to the gravitational acceleration $g$, the density scale height $H=-(\partial\ln{\rho}/\partial\ln{z})^{-1}$, and the sound speed $c_s$ by,
\begin{equation}
    N^2=\frac{g}{H}-\frac{g^2}{c_s^2},
\end{equation}
and
\begin{equation}
    \omega_c^2=\frac{c_s^2}{4H^2}\bigg(1-2\frac{\partial H}{\partial z}\bigg) \approx \frac{c_s^2}{4H^2}.
\end{equation}
The linear wave equation, when applied to a non-magnetic plane-stratified atmosphere, produces a simple dispersion relation (dependant on these two frequencies) which identifies regions of $f-k_{\perp}$ space where acoustic waves can propagate, acoustic-gravity waves can propagate, and where the waves are evanescent \citep[see][and references therein]{srivastava2021chromospheric}. Summarised as,
\begin{equation}
    \omega^2 \gtreqless \frac{1}{2}\bigg(c_s^2k_{\perp}^2+\omega_c^2\pm\sqrt{(c_s^2 k_{\perp}^2+\omega_c^2)^2-4c_s^2N^2k_{\perp}^2} \bigg),
\end{equation}
where the acoustic wave frequencies are > with the + sign, and the gravity wave frequencies are < with the - sign. In a more realistic atmosphere, this picture is complicated by structuring in density, and the presence of the magnetic field. The presence of strong magnetic fields can provide pathways for acoustic waves to propagate up into the chromosphere and corona \citep[magnetic ``pinholes'' or ``windows'';][]{spruit1981motion, jefferies2006magnetoacoustic}. However to first order, the simplified picture holds and is a useful tool. By averaging the properties of the Z = 0 Mm layer, the cut-off frequencies are estimated and the evanescent region is over-plotted in Figure \ref{figureVzFFT}. In the solar atmosphere, the acoustic cut-off frequency is estimated to be around 5mHz \citep{vigeesh2017internal}, which effectively traps the acoustic waves excited by the convective motions below the photosphere. These waves reflect and resonate inside the solar convection zone, forming the well known p-mode oscillations \citep{ulrich1970five}. 

Our simulation does not encompass enough of the convection zone for the classical 5-minute p-mode oscillations to develop. However, as the computational domain is small, and a pressure node is set at the lower boundary, the acoustic waves generated by the convection can create quasi-standing modes with large amplitudes in the upper-convection zone . These pressure-modes \citep[discussed in][]{fleck2021acoustic}, are clearly distinguishable in Figure \ref{figureVzFFT}. Due to the finite depth of the upper-convection zone ($\sim 2$ Mm), and the travel time of acoustic waves ($\sim10$ km/s), positive interference is created at similar frequencies to the classical p-modes observed on the Sun ($\sim300$ s). Along with the pressure-modes (henceforth referred to as ``box-modes''; to indicate their origin), the convective flows shape the velocity PSDs, primarily at low frequencies. The dominant frequencies in the low-corona for both the velocity components correspond to the vertical convective motions, and the box-modes. Comparing the PSD at 4 Mm and 12 Mm in Figure \ref{figureVzFFT}, there is little change in the low-corona for vertical fluctuations. In contrast, Figure \ref{figureVxyFFT} shows the horizontal fluctuations generally decrease in amplitude with increasing height in the domain, except for the highest frequencies. 

\begin{figure*}[t!]
 \centering
  \includegraphics[trim=0.4cm 0.3cm 0.5cm 0.3cm, width=\textwidth]{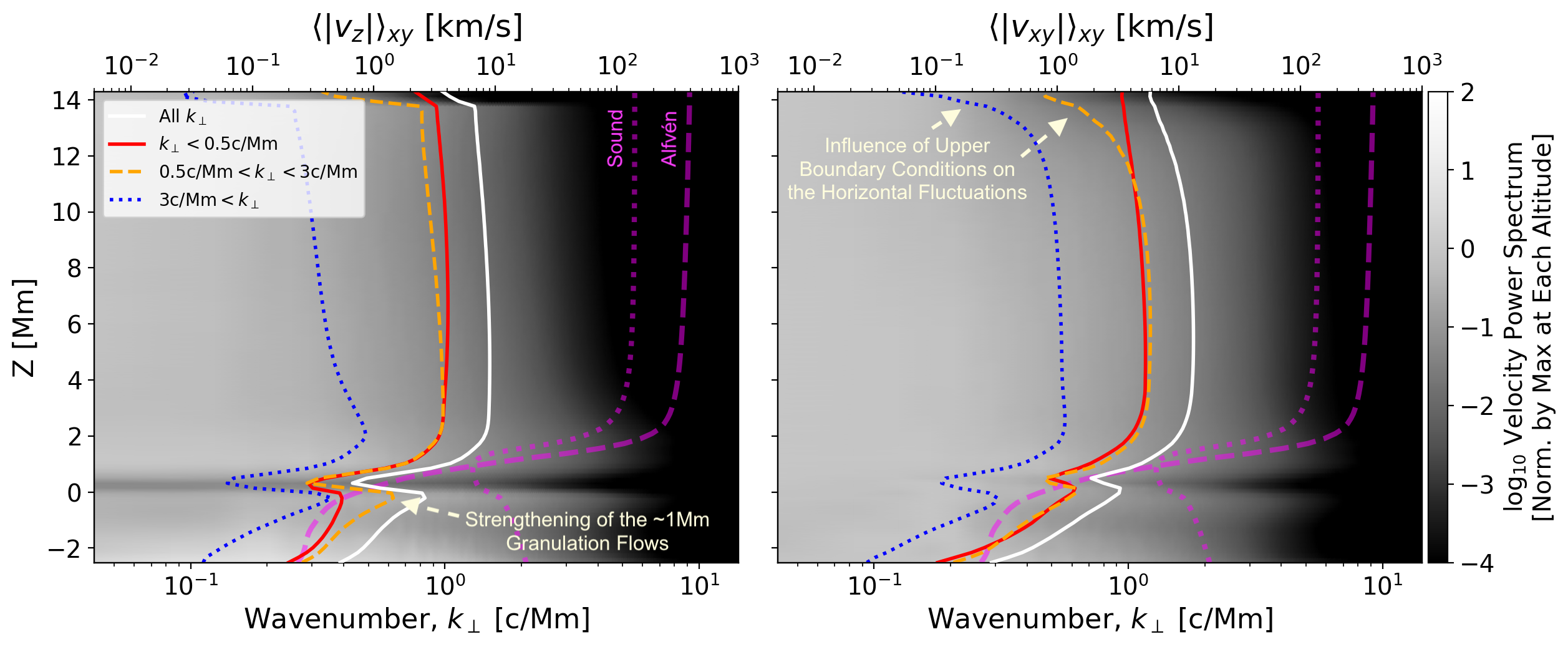}
   \caption{Wavenumber power spectrum of the vertical (left) and horizontal (right) components of velocity, versus height (in both cases normalised by the maximum horizontal PS at each height). The PS of the vertical and horizontal flows become almost identical around Z = 1 - 1.5 Mm likely as a result of mode-conversion and/or magnetosonic shocks at the $v_A/c_s=1$ surface, which allows for wave-energy to pass between longitudinal and transverse oscillations. The horizontally-averaged velocity amplitude in overploted with a solid white line for each component. The amplitudes of the filtered velocity components from three wavenumber ranges, are similarly over-plotted with coloured lines. The time-averaged sound and Alfv\'en speeds versus height are also shown with magenta dotted and dashed lines, respectively.}
   \label{figureKcmap}
\end{figure*}

\begin{figure*}[t!]
 \centering
  \includegraphics[trim=0.4cm 0.3cm 0.5cm 0.3cm, width=\textwidth]{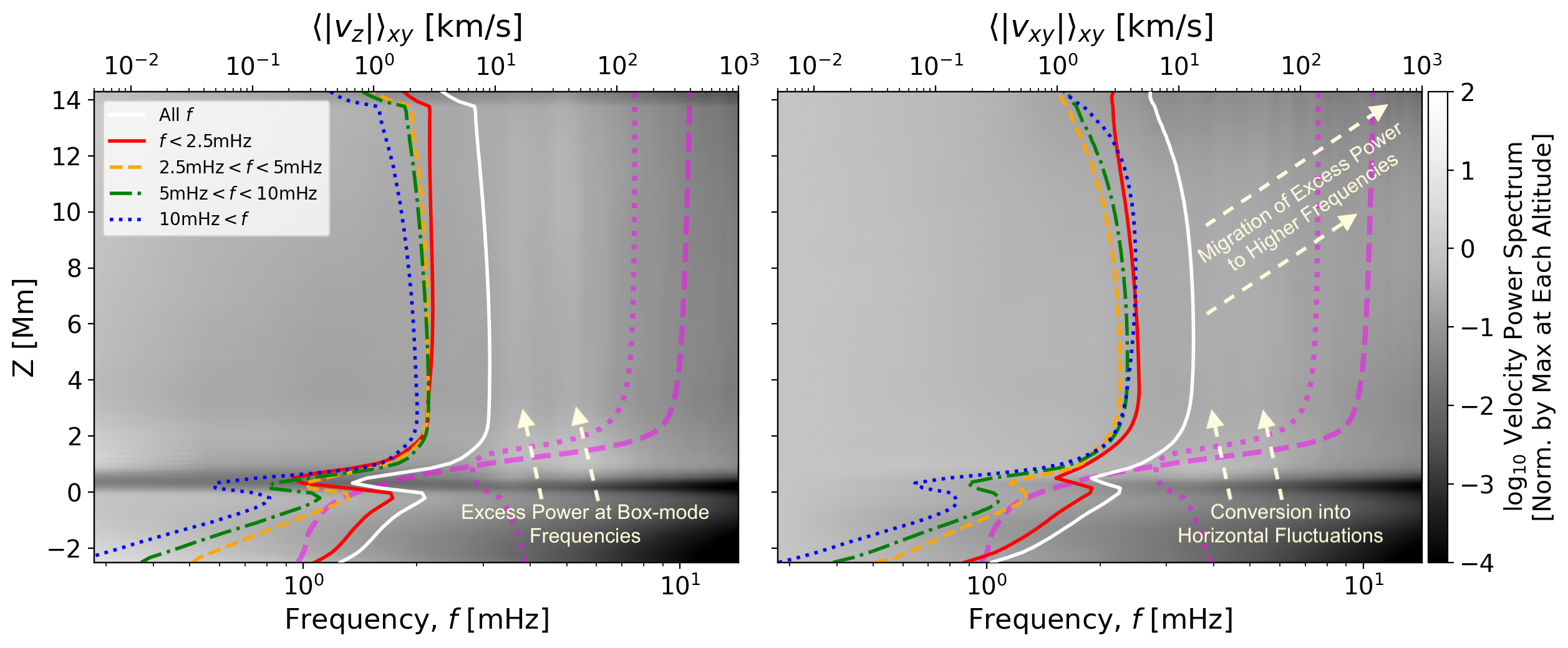}
   \caption{Same as Figure \ref{figureKcmap}, but now showing the frequency power spectrum versus height. Peaks in the PS can be found at 5 - 10 mHz which correspond to the box-modes frequencies (see Figure \ref{figureVzFFT}). In the vertical flow speed, the box-modes are able to survive into the low-corona by travelling along the network of magnetic funnel structures. A similar, but less pronounced, enhancement in the PS is visible in the horizontal flow speed, which moves to progressively higher frequencies with height. }
   \label{figureFcmap}
\end{figure*}

To illustrate further the evolution of the velocity components with altitude, we calculate the Power Spectrum (PS) from the PSD, at various wavenumbers (given in units of cycles/Mm) and frequencies (given in units of mHz), i.e., 
\begin{equation}
    \text{PS}(k_{\perp})=\sum_{\omega}\text{PSD}(k_{\perp},\omega)\delta \omega,
\end{equation}
and,
\begin{equation}
    \text{PS}(\omega)=\sum_{k_{\perp}}\text{PSD}(k_{\perp},\omega)\delta k_{\perp}.
\end{equation}
Figures \ref{figureKcmap} and \ref{figureFcmap} show the resulting PS versus height in the simulation domain. The horizontally-averaged amplitude of the velocity components $v_z$ and $v_{xy}$, versus height in the domain, are over-plotted with solid white lines. The average sound and Alfv\'en speeds are shown for comparison with magneta dotted and dashed lines, respectively. To make clear any frequency/wavenumber-dependent changes in the velocity amplitudes, the horizontally-averaged velocities of three wavenumber ranges: less than 0.5 cycles/Mm (solid red), 0.5 - 3 cycles/Mm (dashed orange) and greater than 3 cycles/Mm (dotted blue), along with four frequency ranges: less than 2.5 mHz (solid red), 2.5 - 5 mHz (dashed orange), 5 - 10 mHz (dash-dot green), and greater than 10 mHz (dotted blue), are also displayed in Figures \ref{figureKcmap} and \ref{figureFcmap}. The wavenumber ranges are an arbitrary choice, unlike the frequency ranges which are chosen to match those used in \citet{shoda2018high}. The spatially-reconstructed velocity components, from which the horizontal averages were calculated are discussed in Appendix \ref{reconstruction}. Horizontal cuts of the frequency-filtered velocities $v_z$ and $v_{xy}$ are shown at the same four heights used in Figures \ref{figureVzFFT} and \ref{figureVxyFFT}, at $t=40$~mins in Figures \ref{figureFilteringVertical2},  \ref{figureFilteringHorizontal2}, \ref{figureFilteringVertical} and \ref{figureFilteringHorizontal}.   

As plasma rises to the top of the upper-convection zone, there is a progression of energy from large to small scales in both velocity components. Near the photosphere, the large-scale convective motions breakdown into the smaller granulation flows. In Figure \ref{figureKcmap}, a shift in the amplitude of the vertical flow towards wavenumbers at the scale of the granulation is observed (see the $k_{\perp}$ = 0.5 - 3 cycles/Mm range). The recirculating plasma at the photosphere also generates increased power in the horizontal flow. Just above the photosphere, there is a decrease in the strength of the PS, as the movement of chromospheric plasma is restricted inside small regions of closed network field. The subsequent increase in strength of the PS towards the top of the chromosphere/low-corona results from three main factors; 1) the decreasing density allows for larger velocities with the same kinetic energy, 2) the motion of the magnetic field is transferred to the plasma as the plasma-$\beta$ regime has reversed, and 3) the geometrical expansion of the magnetic funnels in regions of the chromosphere where the Mach number of the flow is greater than one enhances existing motions. 

In Figure \ref{figureFcmap}, there is a clear enhancement of the vertical flows in the low-corona around 4 - 6 mHz. This corresponds to the range of box-mode frequencies at the photosphere, that appear to resonate in the atmosphere above. A smaller enhancement in this frequency range is visible in the horizontal flow. Unlike the vertical fluctuations, which remain in the same range throughout the low-corona, the horizontal fluctuations undergo dissipation and spread to higher frequencies with altitude. This is likely due to the dissipation of the torsional oscillations in the magnetic funnels. The high frequency horizontal fluctuations continue to gain amplitude until around Z = 10 Mm. The lower frequency modes, show a corresponding decrease as energy is being dissipated into the higher frequencies. These fluctuations have the largest amplitude of the four reconstructed frequency ranges at Z = 8 - 12 Mm, after which they are influenced by the upper boundary condition. In fact, the upper boundary conditions influence all of the flow components (most significantly high frequencies and wavenumbers), though in most cases this is limited to above Z = 14 Mm. 

\begin{figure*}[t!]
 \centering
  \includegraphics[trim=0.0cm 0.0cm 0.0cm 0.0cm,clip, width=\textwidth]{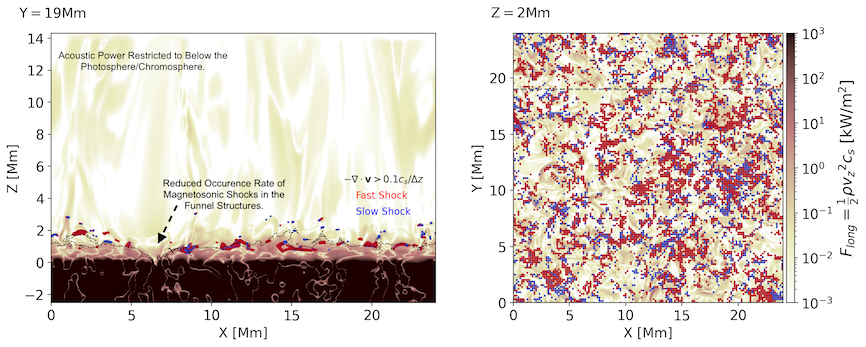}
   \caption{Longitudinal (acoustic) wave-energy flux at $t=40$~mins for a vertical cut through the domain at Y = 19 Mm, and horizontal cut at Z = 2 Mm. Fast and slow magnetosonic shocks, which are identified through a threshold on $\nabla\cdot{\bf v}$, are indicated in red and green colours, respectively. For the horizontal slice, shocks are highlighted at a range of heights throughout the chromosphere (not just from Z = 2 Mm). The plasma $\beta$ equal to one surface is highlighted with a thin dashed black line.}
   \label{figureFilteringVerticalLong}
\end{figure*}

\begin{figure*}[t!]
 \centering
  \includegraphics[trim=0.0cm 0.0cm 0.0cm 0.0cm,clip, width=\textwidth]{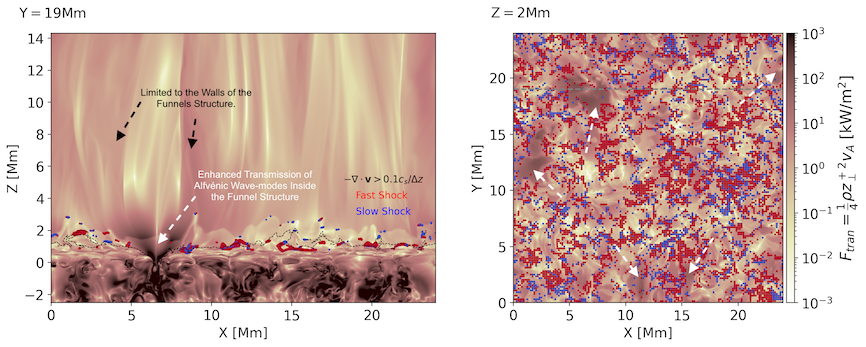}
   \caption{Same as Figure \ref{figureFilteringVerticalLong}, but now for the transverse (Alfv\'en) waves.}
   \label{figureFilteringVerticalTrans}
\end{figure*}

\begin{figure*}
 \centering
  \includegraphics[trim=0cm 0cm 0cm 1cm, clip, width=0.95\textwidth]{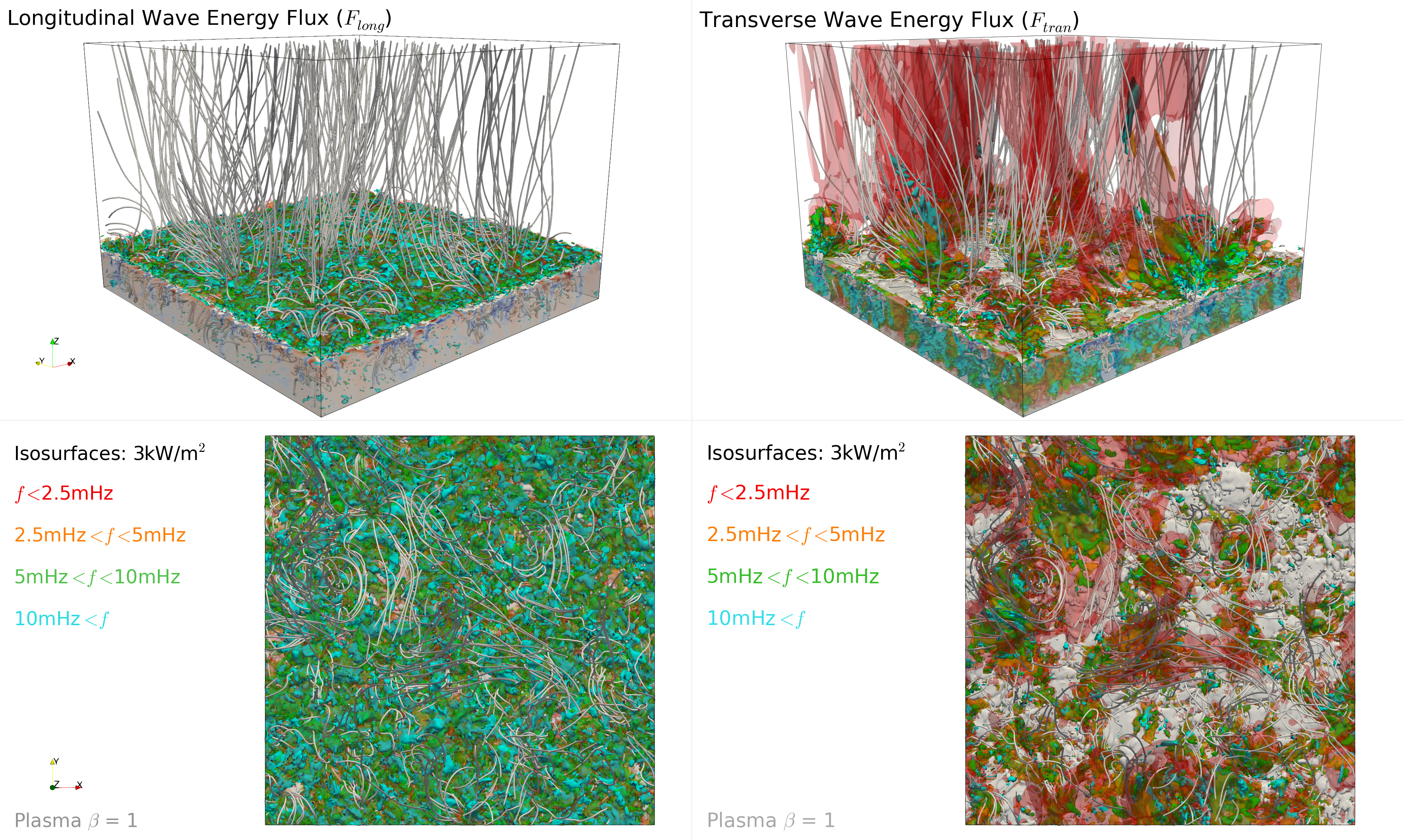}
   \caption{A 3D visualisation of the $F_{long}$ and $F_{tran}$ wave-energy fluxes at $t=40$~mins. The magnetic field structure previously presented in Figure \ref{figureFieldStructure}, is now shown in grey. Semi-opaque isosurfaces highlight the wave-energy fluxes at 3 kWm$^{-2}$ in each reconstructed frequency range; less than 2.5 mHz (red), between 2.5 mHz and 5 mHz (orange), between 5 mHz and 10 mHz (green), and greater than 10 mHz (cyan). The white opaque surface represents plasma $\beta$ equal to one.}
   \label{figure3DFlongFtran}
\end{figure*}

Fluctuations travelling from the photosphere into the chromosphere should be significantly suppressed/damped by the evanescence of the acoustic modes. Yet the amplitude of the velocity fluctuations increases above the chromosphere. Using a simplified MHD model, \citet{snow2018magnetic} showed that vortex/funnel structures can act as wave-guides, and that the interaction of twisted funnels can lead to an enhanced energy transfer from the photosphere. In our simulation, the transmission/reflection of waves through the chromosphere, transition region, and into the corona is difficult to follow due to the many geometrical and dissipative processes that influence their propagation. However, the typical amplitudes of the longitudinal (assumed acoustic) and transverse (assumed Alfv\'en) waves in the computational domain can be readily computed. Their wave-energy fluxes $F_{wave}$ are given by the product of the energy in a given wave mode $E_{wave}$ and its corresponding wave speed $v_{wave}$. This methodology follows closely that used in \citet{shoda2018high}. It is important to note that $F_{wave}$ is not a conserved quantity here, as there are significant vertical flows in the domain, in addition to the wave-energy flux being time-dependent. Regardless, a Fourier analysis is performed, as done previously for the velocity components, now with the following wave-energy fluxes. For longitudinal acoustic waves travelling vertically through the domain, their wave-energy flux is defined as,
\begin{equation}
    F_{long}=\frac{1}{2}\rho {v_z}^2 c_s.
\end{equation}
For the transverse waves, Alfv\'enic fluctuations are considered, and so the Elsasser variables are defined as,
\begin{equation}
    {\bf z^{\pm}_{\perp}} = {\bf v_{xy}} - \text{sign}(B_{z})\frac{{\bf B_{xy}}}{\sqrt{\mu_0 \rho}}.
\end{equation}
Note the Elsasser variables are not characteristic variables in a compressible plasma (discussed in Section \ref{comparison}), as they are defined for an incompressible plasma \citep{marsch1987ideal}. However, they are expected to produce a reasonable assessment of the amplitude of transverse Alfv\'enic fluctuations \citep{magyar2019nature}. Similarly to $F_{long}$, the transverse wave-energy flux is defined as,
\begin{equation}
    F_{tran}=\frac{1}{4}\rho {z^+_{\perp}}^2 v_A.
\end{equation}
It can be shown that $F_{tran}$ relates directly to the field shaking/shearing Poynting flux $S_{z,fs}$ by,
\begin{equation}
    S_{z,fs}=\frac{1}{4}\rho {z^+_{\perp}}^2 v_A - \frac{1}{4}\rho {z^-_{\perp}}^2 v_A,
    \label{SZFStoALFVEN}
\end{equation}
where the two terms correspond to the upwards and downwards propagating Alfv\'enic waves, respectively.

Figures \ref{figureFilteringVerticalLong} and \ref{figureFilteringVerticalTrans}, present vertical and horizonatal cuts of the wave-energy fluxes in the computation domain. The upper-convective zone is a reservoir of acoustic wave-energy however, as previously discussed, the acoustic waves are significantly damped due to the evanescent regions of the atmosphere. Acoustic waves with frequencies above the acoustic cut-off frequency, could propagate into the low-corona, however the convective motions produce predominantly low frequency waves. Magnetosonic waves that succeed to enter the chromosphere typically produce shocks, which are also highlighted in Figures \ref{figureFilteringVerticalLong} and \ref{figureFilteringVerticalTrans} (discussed in Section \ref{shocks}). The location of the vertical cut in Figures \ref{figureFilteringVerticalLong} and \ref{figureFilteringVerticalTrans} is chosen so that it passes through a magnetic funnel (shown in detail in Figure \ref{figureSwirling}). This magnetic funnel can be seen to act as a gateway to the low-corona for transverse waves, which are driven by photospheric motions. From the Fourier transforms of $F_{long}$ and $F_{tran}$, the spatial variation of each wave-energy flux is reconstructed in four different frequency ranges. A 3D visualisation of the wave-energy fluxes is shown in Figure \ref{figure3DFlongFtran}. The limited vertical extent of the longitudinal wave-energy is again easily observed, along with the enhanced transverse wave-energy flux inside the magnetic funnels (in areas without twisted magnetic funnels, the transverse wave-energy is typically confined below the plasma $\beta=1$ surface).

The transverse wave-energy flux is most often in the form of torsional oscillations in the braided magnetic funnel structure. At higher frequencies these oscillations are limited to the outer edges of the funnels, and are subsequently smaller-scale. The lowest frequencies recover the twisting/undulating motion of the entire low-corona, driven by the persistent buffeting/shuffling of MCs around the intergranular lanes (where most of the open magnetic field is rooted). The majority of transverse wave-energy reaches the top of the domain in the low frequency oscillations. The lowest frequencies ($f<2.5$ mHz) have $F_{tran}$ around a factor of ten larger than the higher frequency ranges at Z = 14Mm. 

\begin{figure*}[t!]
 \centering
  \includegraphics[trim=0.4cm 0.3cm 0.5cm 0.3cm, width=\textwidth]{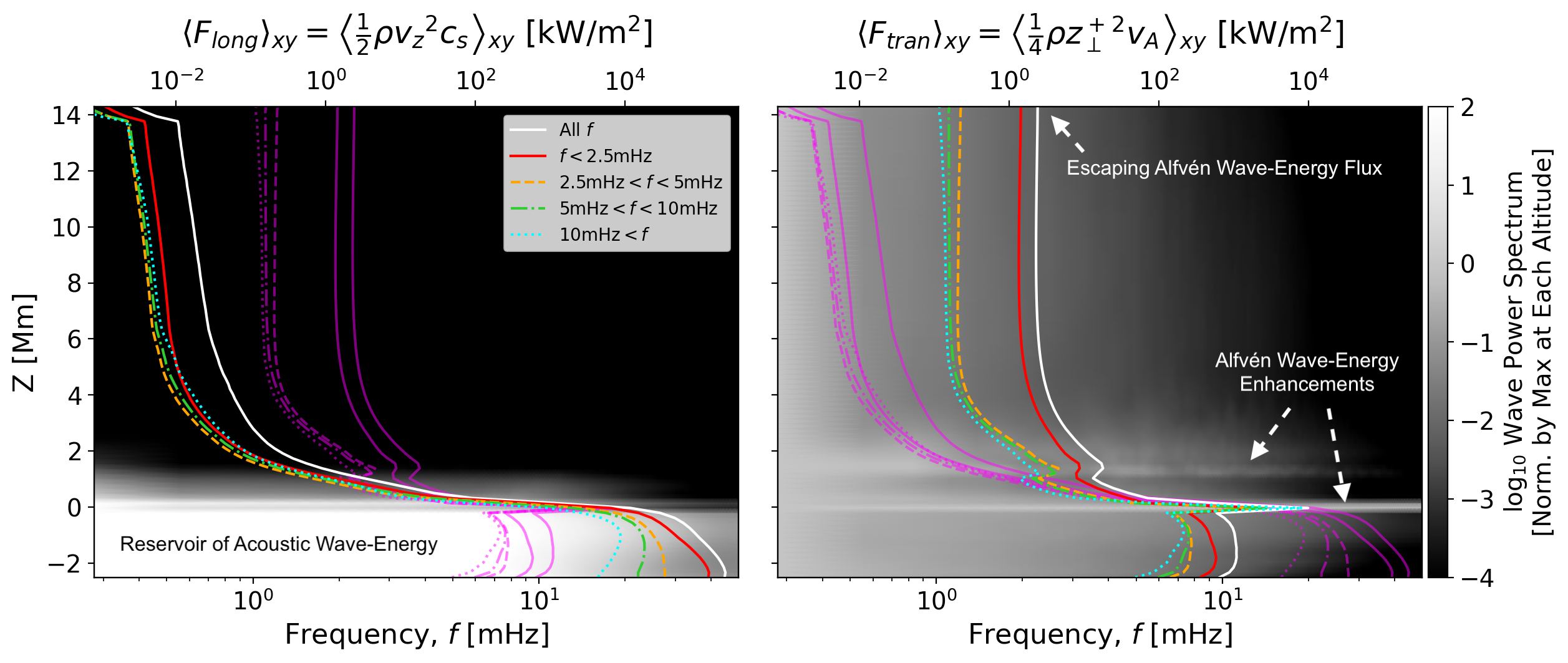}
   \caption{Frequency power spectrum of the longitudinal (acoustic) wave-energy flux $F_{long}$ (left) and the transverse (Alfv\'en) wave-energy flux $F_{tran}$ (right), versus height (in both cases normalised by the maximum transverse PS at each height). Coloured lines are over-plotted showing the average energy flux versus height for four frequency ranges and their total. Grey lines show the other wave flux averages for comparison. The upper-convection zone is a source of both longitudinal and transverse waves, with the longitudinal waves having an energy flux that is 2 - 3 orders of magnitude larger than the transverse waves. However, the longitudinal wave-energy flux decays exponentially with height. Such that in the low-corona, the relative balance of energies is reversed. }
   \label{figureLongVsTran}
\end{figure*}

Figure \ref{figureLongVsTran} shows the PS of the wave-energy fluxes in frequency space versus height in the domain. The average amplitude of the wave-energy fluxes versus height are over-plotted along with the averages of four reconstructed frequency ranges. As observed in Figure \ref{figureFilteringVerticalLong}, there is a large amount of energy in the acoustic waves/convective motions in the upper-convection zone. However, this wave-energy is largely confined below the photosphere, such that the average energy flux versus height decays rapidly. The Alfv\'enic waves have far less energy in the upper-convection zone, but are enhanced at two altitudes: the photosphere (Z = 0 Mm) and the $c_s=v_A$ surface (Z$\approx$ 1.5 Mm). At the photosphere, the plasma and magnetic field strongly coupled, with MCs being directly driven by convective motions. Power in the Alfv\'en waves therefore rises through the photosphere, approaching equivalence with the power in the acoustic waves. This effect is seen in all of the frequency ranges (coloured lines). After this, the transverse fluctuations are similarly damped like the acoustic waves in the chromosphere. However, around Z = 1.5 Mm the transverse wave-energy flux is once again enhanced. At this height the sound speed and Alfv\'en speeds are equivalent, so mode-conversion \citep{schunker2006magnetic,cally2007three} may take place (note that the decay of the acoustic wave-energy flux becomes stronger here). Magnetosonic shocks also form at this height, driven by both upward propagating waves and previously ejected plasma falling back down to the chromosphere. These shocks may dissipate energy into Alfv\'en waves as they shake the surrounding field lines. In addition to these enhancements, the funnel network helps to reduce the dissipation/reflection of the transverse wave-energy as inside the gradients in density and local Alfv\'en speed are smaller. Frequently, Alfv\'enic fluctuations have their amplitudes increased by the expansion of the funnels with Mach number greater than one flows. Once above the chromosphere, Alfv\'enic waves propagate with little dissipation in the magnetically dominated low-corona. The amplitude of wave-energy flux that propagates up into the low-corona, and would subsequently enter the solar wind, is quantitatively discussed in Section 4.

\subsection{Magnetosonic Shocks}\label{shocks}
Due to the sharp density gradients in the solar atmosphere, MHD waves can transform into magnetosonic shocks \citep{delmont2011parameter}. The location of magnetosonic shocks inside the \simulation{} simulation are identified by a criterion on the extreme negative values of $\nabla\cdot{\bf v}$, as done previously by \citep[][and reference therein]{wang2020simulation}. The frequency distribution of $\nabla\cdot{\bf v}$ around zero is symmetric for linearly propagating waves in the domain. The presence of shocks creates an asymmetry in the distribution by increasing the frequency of extreme negative values. Thus, by selecting the grid cells responsible for the asymmetric tail of the $\nabla\cdot{\bf v}$ distribution, the location of shocks in the domain can be found. However, there will be some contamination via linear waves with large amplitudes. The criterion for shock selection is given,
\begin{equation}
    -\nabla\cdot{\bf v}>\xi c_s/dz,
\end{equation}
where $\xi$ is a free parameter that is used to set the threshold for identification, based on the sound speed and the numerical grid spacing. The vertical grid spacing $dz$ is used here for two reasons, 1) the shocks are typically orientated towards the vertical, and 2) the vertical grid spacing changes with altitude so this effectively accounts for that in the selection criterion. As done in the Appendix of \citet{wang2020simulation}, we manually select a value of $\xi=0.1$ which avoids the majority of the linear compressions (based on the extent of the positive distribution).

Once the shocks are identified, the shock direction is derived based on the local pressure gradient. To identify the type of shock, the magnetic pressure gradient is compared to the plasma pressure gradient over the shock. If the two pressure gradients are aligned, the shock is labelled as a fast-magnetosonic shock. If the two pressure gradients are oppositely directed, the shock is labelled as a slow-magnetosonic shock. The location of detected shocks are displayed in Figures \ref{figureFilteringVerticalLong} and \ref{figureFilteringVerticalTrans}. It is well known that shocks play a vital role in heating the chromosphere \citep{carlsson1992non, carlsson1997formation}. Thus it may be possible to constrain their heating rates with the \simulation{} simulation, however a quantitative analysis is left for future work. 

Shocks are produced throughout the chromosphere in the simulation, as waves encounter the rapidly decreasing density gradient and shock. Longitudinal waves can avoid this fate by undergoing mode-conversion into a transverse wave mode, which can propagate more freely in the low-corona. The shocks detected inside the chromosphere of the simulation are more typically fast-magnetosonic shocks, at an altitude of around 1.5-2Mm. The formation of shocks appears to be suppressed inside the twisted magnetic funnels structures (see arrows in horizontal cut of Figure \ref{figureFilteringVerticalTrans}). This may arise from the locally enhanced wave transmission, reducing the likely-hood of wave-energy dissipation into shocks. At times when the funnel structures undergo SEs, chromospheric plasma is typically launched into the low-corona. This can create a small number of slow-magnetosonic shocks higher up in the solar atmosphere (around 3-6Mm above the photosphere). This is unlikely to represent a large amount of energy (especially compared with the Poynting flux enhancements), but illustrates that SEs may trigger further mechanisms of energy dissipation higher up in the solar atmosphere.

\section{Implications for the Solar Wind}
This study aims to establish connections between the small-scale dynamics observed in the realistic solar atmosphere produced by \bifrost{}, and the heating/acceleration of the solar wind. In this Section, the simulation results are discussed in the context of recent observations by Parker Solar Probe (PSP), which is currently measuring the solar wind closer to the solar surface than ever before \citep{PhysRevLett.127.255101}. Following this, we argue that high-resolution imaging of the solar chromosphere \citep[see review of][and reference therein]{carlsson2019new} may be able to validate some of our conclusions. 

\begin{figure*}[t!]
 \centering
    \includegraphics[trim=0.0cm 0.0cm 0.0cm 0cm,clip, width=\textwidth]{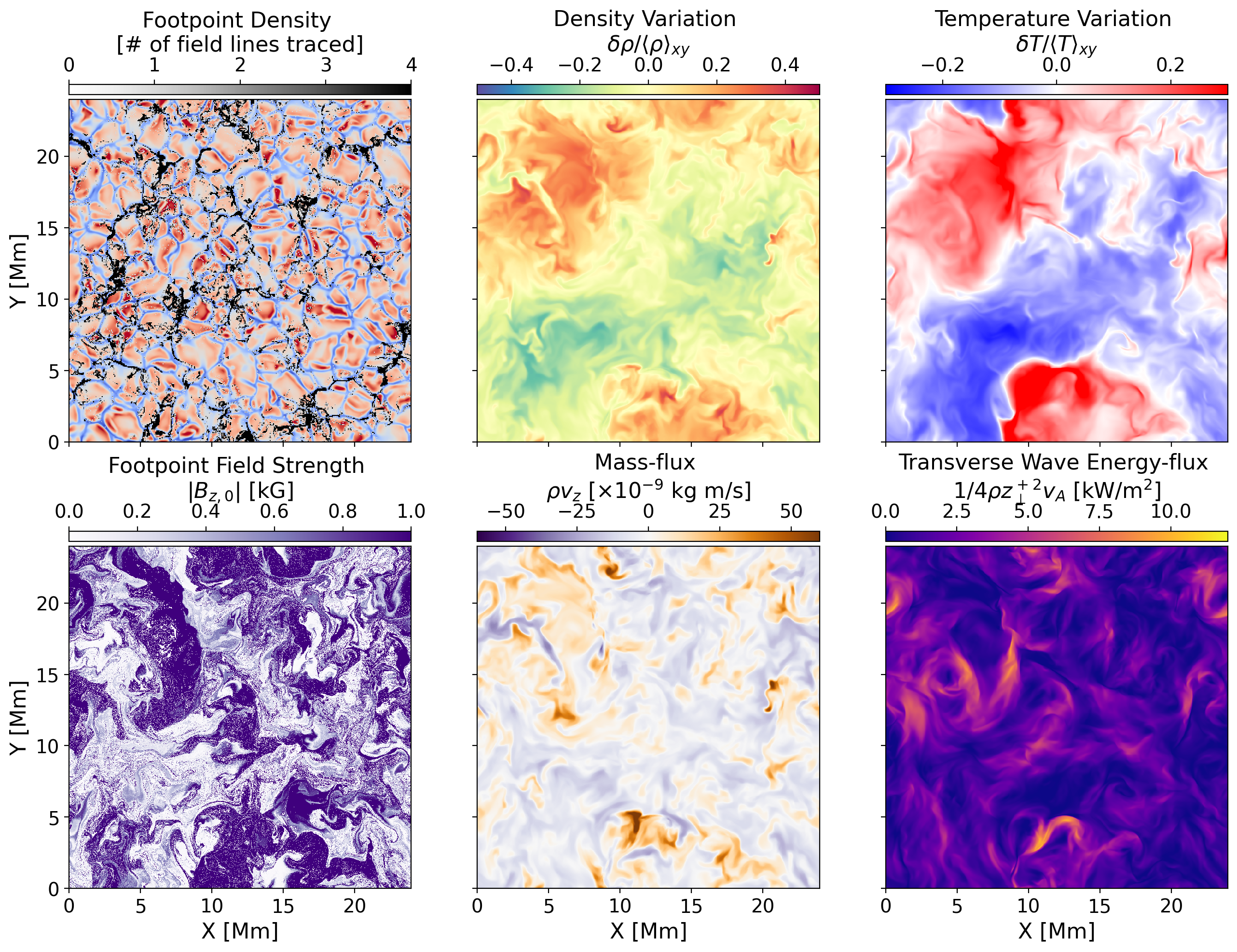} 
   \caption{Top Left: Location of the open magnetic field lines at the photosphere (traced down from Z = 12 Mm), with the convection pattern shown bellow. Bottom Left: Open magnetic field at Z = 12 Mm coloured by their photospheric field strength. Top Middle and Right: Density and Temperature variation with respect to the horizontal average at Z = 12 Mm. Bottom Middle: Mass flux at Z = 12 Mm. Bottom Right: Transverse wave-energy-flux at Z = 12 Mm. Each panel is taken at $t=40$~mins. The contrast in density, temperature and upward wave-energy is evident, resulting from the different MC field strengths at the base of the braided magnetic field structures.}
   \label{figureQfactor}
\end{figure*}

\subsection{Generation of Flux Ropes}\label{fluxropes}
The nature of the Sun's magnetic field, and subsequently that of the interplanetary magnetic field, has been shown to be comprised of smaller twisted strands \citep[this is visibly apparent in the context of coronal loops][]{williams2020high}. The term \textit{flux rope} (or \textit{flux tube}) is used throughout the heliophysics literature to describe these smaller structures. Our simulation results show that the magnetic field of the low-corona can be readily braided into flux rope-like strands by the motion of MCs along the intergranular lanes. The magnetic field is braided in such a way that there are naturally differing degrees of twist between strands in the flux rope, which is necessary for their stability \citep{wilson1977elementary}. Assuming that the simulation captures the smallest meaningful scales of this braiding, then the diameter of flux robes should depend on the scale of the granulation pattern (more specifically the spatial distribution of intergranular lanes in which the magnetic flux is transported) and the strength of any dissipative processes that can untangle the field. 

Ohmic heating and viscous dissipation are largest at the borders of the twisted magnetic funnels, however their absolute strengths are relatively small in our simulation ($<$3 mW/m$^2$) when compared with the average kinetic, and magnetic energy fluxes ($\sim$5 W/m$^2$, and $\sim$0.6 kWm$^{-2}$, respectively) above 6 Mm in altitude. Therefore, the twisted structures are subject to negligible dissipation in the low-corona after their formation, and are stable. In reality, these dissipative terms may differ in strength to those used in our simulation (perhaps due to non-ideal processes). By increasing the dissipation, we might expect to generate smaller flux ropes, as the field can more easily untwist. We remind the reader that the \bifrost{} simulations operate with a split-diffusive operator for the diffusive terms in the MHD equations. Further studies that can effectively control these parameters may be needed to quantify exactly how large these braided structures can become. According to our computation, the average flux rope weaves together around 2 - 3 photospheric sources (MCs) such that in the low-corona, they have a combined diameter of 5 - 15 Mm. Isolated magnetic funnels that are stressed into twisted configurations (without the addition of neighbouring flux) have most-often diameters of 4 - 6 Mm. These isolated magnetic funnels are often subject to SEs, and so this sets the diameter of Poynting flux enhancements. However, it is difficult to quantify how much these sizes are influenced by the size of the computational domain.

\subsection{Structured Outflow}\label{structure}

PSP is now routinely sampling the solar wind below 0.25 au, with the ultimate goal of making in-situ measurements as close as $10R_{\odot}$ \citep[][]{fox2016solar}. Such unprecedented access to the emerging solar wind has given authors the opportunity to make connections between the structure at the base of the wind and the structure observed in-situ. This is especially true for the origins of reversals in the polarity of the solar wind magnetic field, referred to as \textit{switchbacks} \citep{owens2018generation, kasper2019alfvenic, mozer2020switchbacks, mcmanus2020cross, macneil2020evolution}. Switchbacks are frequently observed in patches \citep{de2020switchbacks}, that have been inferred to correspond to the scales of super granulation at the solar surface \citep[see][and references therein]{fargette2021characteristic}. \citet{bale2021solar} suggest that the spatial variation of switchback patches corresponds to how the solar wind magnetic field emerges from the network of magnetic funnels rooted in the photosphere. Thus the results from the \simulation{} simulation may directly support or invalidate this hypothesis. 

As discussed throughout this study, in our simulation, the outflowing material is structured by the underlying network of magnetic funnels. The left panels of Figure \ref{figureQfactor} present a snapshot of the footpoints of the open magnetic field in the photosphere (on top of the convection pattern), along with the variation of the footpoint field strength in the open magnetic field (when traced down from Z = 12 Mm). The variation in plasma density and temperature are also depicted at an altitude of Z = 12 Mm. Along with the mass flux and transverse wave-energy flux at Z = 12 Mm. \citet{bale2021solar} report that the solar wind inside the switchback patches is faster, and hotter than the background solar wind. Our simulation does indeed produce a temperature contrast between elements of the funnel network however, instead of a variation between the centre of the funnels and their edges \citep[as suggested by][]{bale2021solar}, this corresponds to the inhomogeneous braiding of the coronal field (and the subsequent releases of energy in SEs). There is a large variation in the photospheric magnetic field strength of neighbouring funnels in the low-corona, with SEs typically occurring in those with the strongest magnetic fields (identified by MCs). An intermittent mass-flux is driven into the low-corona above these MCs, which is accompanied by an enhanced Poynting flux (due to the SEs). Therefore, the low-corona above MCs has an increased temperature, and density (both $\sim25\%$ larger at Z = 12 Mm) with respect to the average coronal plasma. The enhanced Poynting flux also indicates a more effective transmission of Alfv\'en wave-energy, which may ultimately feed the turbulent-formation of switchbacks in the solar wind.

Given the limited vertical extent of our simulation, it is difficult to say how the patchy heating of the coronal funnel network by ohmic heating and SEs will influence the formation of the solar wind above, or if this contrast in properties at the base of the solar wind is enough to drive the variations observed in-situ. It is unclear whether or not these variations will be able to survive out to large distances, without being disrupted by turbulent mixing or dissipative processes. Work from \citet{borovsky2021solar} appears to suggest that various kinds of structures embedded in the solar wind can remain coherent from 1au out to larger distances (despite the turbulent nature of the solar wind), thus the same may be true for variations in the near-Sun environment (though this is purely speculative). The \simulation{} simulation also has a relatively small horizontal scale, in the context of the PSP observations (more consistent with the scales sizes proposed by \citealp{fargette2021characteristic}), thus it is possible that the variations in the solar wind are organised on a larger-scale than is available to us here.

\subsection{Enhancing the Turbulence Generation of Switchbacks}\label{switchbacks}
The exact mechanism(s) that generates switchbacks in the solar wind remains unknown, though there are many theoretical models \citep{squire2020situ, fisk2020global, zank2020origin, mallet2021evolution, schwadron2021switchbacks, drake2021switchbacks, magyar2021could}. Some models favour the formation of switchbacks close to the solar surface as Alfv\'enic fluctuations (often as a result of interchange reconnection), whilst others argue that switchbacks are generated as the solar wind expands into the Heliosphere (via turbulence, shears in velocity, etc). \citet{shoda2021turbulent} find that switchbacks are naturally generated through the non-linear evolution of Alfv\'enic fluctuations with turbulence, which directly links this phenomena to the fundamental heating/acceleration of the wind. However the frequency of occurrence of switchbacks generated through this turbulent process in the model of \citet{shoda2021turbulent} is far lower than measured in-situ by PSP. Shoda et al. (in Prep) further develop this idea, and show that the occurrence rate of switchbacks can be dramatically increased (without significantly increasing the input Poynting flux) by including a coherent torsional driving at the base of the wind (i.e., stirring the field lines in the low-corona). Figure \ref{figureTwistedFluxTubes} and \ref{figureDissipation} show the structure of the Poynting flux in our simulation, which contains a quasi-steady background of torsional Alfv\'en waves along with the stronger SEs. Both of which, given the results from Shoda et al. (in Prep), are expected to increase the rate of turbulent switchback formation. As well as providing a clear example of the torsional driving required to better reconcile the turbulent generation of switchbacks with observations, our simulation also provides a mechanism for clustering the switchbacks into patches due to the varying conditions between neighbouring flux tubes.

\subsection{Chromospheric Swirls}\label{chromoSwirls}
Simulations using similar numerical methods and physics to \bifrost{} \citep[notably the {\fontfamily{lmtt}\selectfont MURaM} code;][]{vogler2005simulations} have commonly found the formation of vortex flows, and/or vortex-like structures in the chromospheric magnetic field \citep{kitiashvili2012dynamics, moll2012vortices,amari2015small,kato2017vortex, rappazzo2019magnetic, yadav2020simulations, silva2020solar}, supporting the existence of these structures in the solar atmosphere. However, in general, these studies have been focused on the chromospheric dynamics, which can produce an observational signature \citep{wedemeyer2009small, battaglia2021alfv}. Such signatures have been observed in high-resolution solar imagery (in observations these structures are often referred to as \textit{swirls}), generally centered over enhancements in the photospheric magnetic field (i.e., MCs), which are embedded in the narrow intergranular lanes. \citet{wedemeyer2012magnetic} were able to show correlated signatures of these swirls at various heights in the solar atmosphere \citep[see also][]{tziotziou2018persistent}, reinforcing again the connection with the stirring motions in this study. Chromospheric swirls have typical diameters of $\sim$2 Mm and lifetimes of around 9 - 10 minutes, in quiet-Sun regions \citep[][]{shetye2019multiwavelength}. The average diameters of SEs in our simulation, and there durations, are comparable to the observations of chromospheric swirls by \cite{shetye2019multiwavelength} and with some observations of solar tornadoes \citep[discussed in;][]{wedemeyer2014plasma}. We propose that some of these chromospheric swirls could be explained by the SEs captured in our simulation, with MCs at their base. This assumes that the Poynting flux carried by the SE is able to heat the chromospheric plasma sufficiently to reproduce the observations, which is left for future works to investigate. 

Our simulated swirls undergo phases of activity with loading and unloading of the magnetic field lines, during which time the chromospheric plasma is heated. This timescale is governed by a number of factors, though the most significant appear to be: 1) the size of the intergranular network, 2) the speed of MCs travelling through the network, and 3) the disspative processes that prevent energy being built up in the twist of the magnetic funnels. Therefore, if chromospheric swirls could be observed in a coronal hole region, the statistics of their sizes, and occurrence frequency, may be used to constrain the injection of heating (and torsional driving) through SEs into the solar wind. 

At present, the authors are not aware of any high-resolution observations of chromospheric swirls (or the lack there of) in coronal hole regions. As coronal holes are often located at the poles of the Sun, this may result from observational bias. The observations of \citet{shetye2019multiwavelength} are taken from quiet-Sun regions, and may not be directly comparable to our simulation results. However, the configuration of the solar atmosphere in areas of quiet-Sun can be very similar to that in areas of coronal holes (just with a closed network of magnetic field that extends higher in altitude). So it may be reasonable to infer that, the presence of chromospheric swirls in the quiet-Sun regions indicates that there is similar braiding of the vertical/open magnetic field by the convective motions below. Thus the observations of chromospheric swirls by \citet{shetye2019multiwavelength} could be an indirect observation of slow solar wind emerging from quiet-Sun regions \citep[slow solar wind sources are discussed in][]{abbo2016slow}. 

\begin{figure*}[t!]
 \centering
  \includegraphics[width=\textwidth]{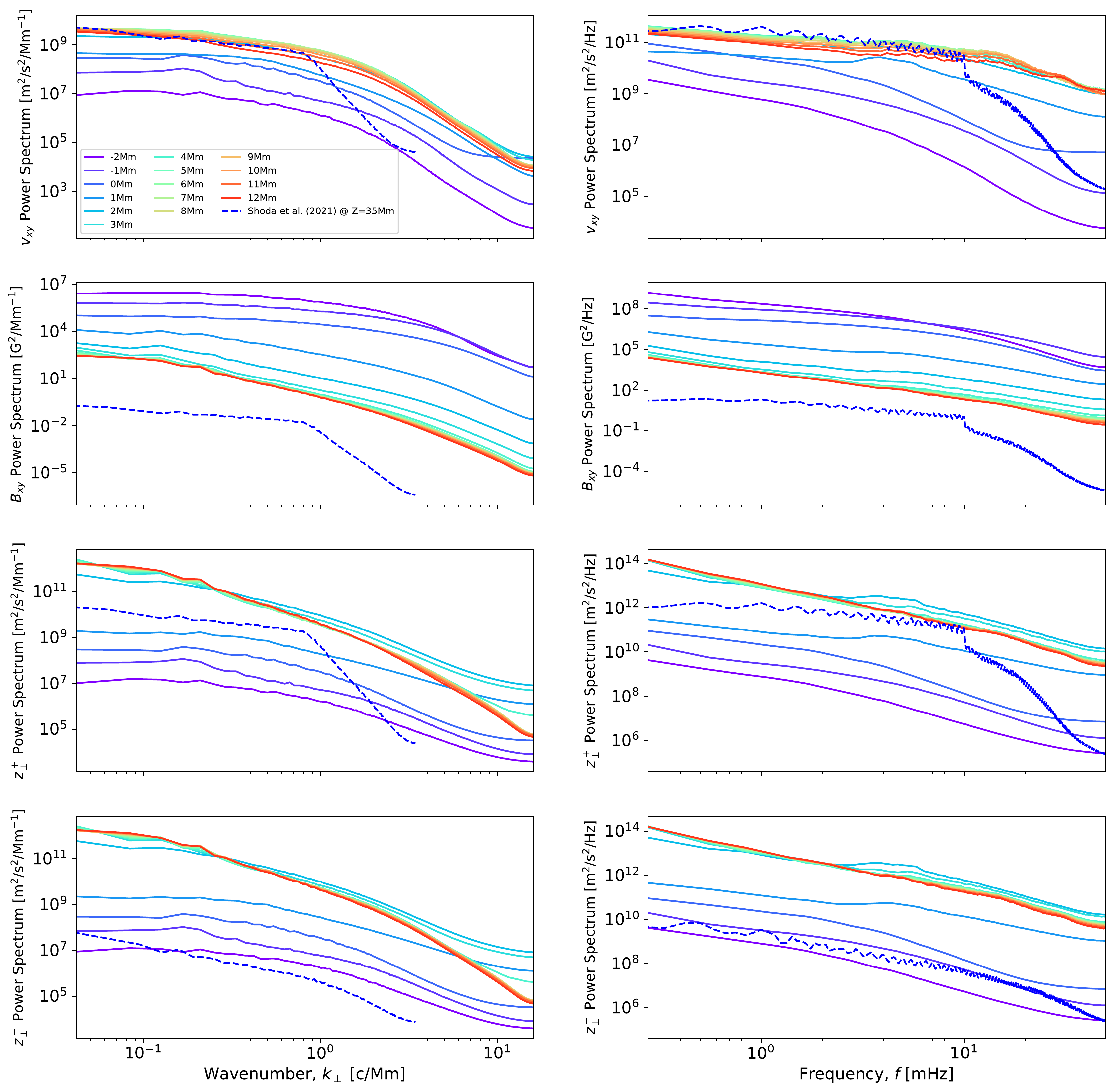}
  \caption{Power spectra versus wavevector and frequency at various heights in the computational domain for the horizontal flow speed, magnetic field strength, and Elsasser variables. The power spectra from the simulation of \citet{shoda2021turbulent} at a height of Z=35Mm are over-plotted for comparison with dashed blue lines.}
  \label{figurePSDcompShoda}
\end{figure*}

\subsection{Implications for Alfv\'en Wave-Driven Wind Models}\label{comparison}

The Alfv\'en wave-turbulence paradigm for heating and accelerating the solar wind has been generally successful at reproducing in-situ observations of the solar wind \citep[see comparison with PSP observations by][]{reville2020role}. However, models that self-consistently capture the propagation and dissipation of the Alfv\'en waves are often limited to 1D \citep{suzuki2011self}, or a 2D/3D wedge configuration \citep{matsumoto2012connecting, shoda2019three, magyar2021three}. In such models, the input wave-energy spectrum is based on the spectrum of observed convective motions and/or the movements of photospheric bright points \citep[see][and reference therein]{cranmer2005generation}, which are assumed to shake the footpoints of the solar wind magnetic field lines. Global models of the solar wind typically favour a parameterised description of this phenomena, which by-passes the need to resolve the scales of the wave propagation and dissipation. This simplified description has a few control parameters, one relating to the input wave-energy flux (discussed in Section \ref{alfvenwaves}), and two-three other parameters describing how the waves dissipate their energy and momentum. 

Simulations performed by the \bifrost{} code are able to follow waves/fluctuations from the upper-convection zone into the low-corona. Understanding how the wave spectrum at the base of the solar wind differs from current models of the solar wind, and in future from different configurations of the \bifrost{} code, is crucial to better constraining models of the solar wind.  In Figure \ref{figurePSDcompShoda}, the PS taken from the \simulation{} simulation at various altitudes in the domain, are compared with the corresponding PS obtained from the simulation of \citet{shoda2021turbulent} at a height of 35Mm (the lowest height with data available). The simulation of \citet{shoda2021turbulent} is performed in a 3D wedge configuration that extends from 1.02 solar radii to 40 solar radii. The horizontal extent of the domain at Z = 35Mm is 31.5 x 31.5 Mm$^2$ (comparable with the 24 x 24 Mm$^2$ of our simulation domain). The magnetic field in the simulation of \citet{shoda2021turbulent} has a strength of 1.1 G at the bottom boundary of Z = 14Mm, which is far weaker than the $\sim5$ G coronal field in our simulation domain. PS of the horizontal flow, the horizontal magnetic field, the positive (outwards) Elsasser variable and the negative (inward) Elsasser variable, are displayed in Figure \ref{figurePSDcompShoda}. Our computational domain does not extend as high as Z = 35 Mm, however the PS evolves slowly with altitude in the low-corona so direct comparison can be justified. 

The amplitude of the horizontal flow at low frequencies and wavenumbers in the \bifrost{} simulation is remarkably similar to that in the model of \citet{shoda2021turbulent}. In our simulation, the horizontal flow amplitude is maximised around Z = 5 Mm. At $f=1$ mHz, both simulations produce a $v_{xy}$ amplitude in the low-corona of $\sim 10$km/s. The two PS depart at higher frequencies, around a few mHz, where our simulation has larger amplitudes.  This is likely a result of the box-modes in the simulation domain which have already been shown to impact the PSD of the velocity components in the low-corona. Higher frequencies are energised by these modes through dissipation. Differences in the amplitude $B_{xy}$ can in-part be explained by the higher coronal density in the \bifrost{} simulation. The influence of a higher density on Alfv\'enic fluctuations is estimated using $B_{xy}\approx\sqrt{4\pi\rho}v_{xy}$, and also the larger vertical field strength $B_z$ in the domain. Using $f=1$ mHz, the field strength amplitude in our simulation is $\sim$3G, compared with $\sim$0.1 G at the same frequency in the model of \citet{shoda2021turbulent}. As the Alfv\'enic relation includes density, and $\rho\approx 5\times 10^{-12}$ kg/cm$^3$ in the low-corona (see Figure \ref{figureAverageProfiles}) which is a factor of ten larger here than in the model of \citet{shoda2021turbulent}. It is expected that for the same value of $v_{xy}$, our $B_{xy}$ should be a factor of $\sqrt{10}\approx3$ larger. The actual factor is around 30, which is significantly larger than predicted by the Alfv\'enic relation alone. The remaining discrepancy is due to the stronger coronal magnetic field strengths in our simulation, compared with that of \citet{shoda2021turbulent}, which are significantly twisted into the horizontal direction. The larger field strengths in our simulation may be explained by the artificial configuration of the simulation domain. As we use a Cartesian box geometry and conserve magnetic flux, the field strength does not decay with altitude. Ultimately, this results in the $z^{+}_{xy}$ Elssasser amplitude at the top of the domain being much larger that the values from \citet{shoda2021turbulent}. 

In Figure \ref{figurePSDcompShoda} the negative Elsasser variable, which in ideal non-compressible MHD describes the inward propagating Alfv\'en waves, is larger than might be expected (in the simulated low-corona $|{\bf z_{\perp}^-}|\approx 0.98|{\bf z_{\perp}^+}|$). As the \bifrost{} code is compressible, the separation of the inward and outward propagating waves into the Elsasser variables is imperfect. In compressible MHD, the outward propagating Alfv\'en waves can appear in the inward variable and vice versa. However this is expected to be a smaller effect than observed here \citep[see discussion of][and references therein]{magyar2019nature}. Therefore it is likely that the open boundary condition at the top of the simulation domain is not perfectly transmitting the Alfv\'en waves out of the domain. This limitation will be discussed further in Section \ref{limitations}. Fundamentally, this means that the inward Elsasser variable from the \simulation{} simulation should be interpreted with caution, but the outward Elsasser variable should still be representative of the upward Alfv\'enic fluctuations.

\subsection{Estimating the Alfv\'en Wave Energy Input}\label{alfvenwaves}

As previously discussed, a parameterised version of Alfv\'en wave-turbulence is used in many global solar wind models \citep[most notably the {\fontfamily{lmtt}\selectfont AWSoM} and {\fontfamily{lmtt}\selectfont wind$\_$predict} models;][]{van2014alfven, hazra2021modeling}. In this description, there are some free parameters that are naturally tuned to get the best results. As the solar wind can be measured in-situ, the optimisation of these parameters is feasible and can provide insight into how much energy is deposited by Alfv\'en waves into the solar wind. However, when the same parameterisations (and even codes) are used in an astrophysical context (e.g. for solar-like stars), the input Alfv\'en wave-energy flux is no longer constrained \citep[see discussion in][]{saikia2020solar}. It is not obvious how the input Alfv\'en wave flux (often denoted $S_A$ in the literature) should change for a young solar-analogue \citep[e.g.,][]{evensberget2021winds}, when compared to the current Sun. To increase the predictive power of these global solar/stellar wind models, it is necessary to understand how changing the magnetic field configuration (and atmospheric structuring) will influence the input Alfv\'en wave-energy into the wind. This requires a better understanding of the progression of energy through the chromosphere and into the corona (under arbitrary conditions), than is currently available from models within the literature. 

The Alfv\'en wave-energy flux $S_A=F_{tran}$ in the low-corona of the \simulation{} simulation is estimated in Section \ref{waves} via equation (\ref{SZFStoALFVEN}). Typically the input energy flux to the wind is expected to be proportional to the field strength itself, and so the input parameter to the solar wind models is typically $S_A/B$ (i.e. the energy flux normalised by the photospheric field strength). Figure \ref{figureLongVsTran} shows the average Alfv{\'e}n wave-energy flux in the low-corona is around 1.5 kWm$^{-2}$ which, when normalised by the average photospheric field strength of 40 G (4 mT), gives a value of $S_A/B=3.8\times 10^{5}$ Wm$^{-2}$T$^{-1}$. This is around a factor of three smaller than the typical input to solar wind models, for example \citet{sokolov2013magnetohydrodynamic} estimate a value of $S_A/B=1.1\times 10^{6}$ Wm$^{-2}$T$^{-1}$. Future work could explore the dependence of the Alfv\'en wave-energy versus photospheric magnetic field strength, for a range of \bifrost{} simulations. However, performing 3D realistic simulations currently remains very challenging and computationally costly. Similar studies have been performed with 1D models \citep[see][]{sakaue2021m}. 

To reduce the need for expensive computations, previous works have used analytical models, that attempt to follow Alfv\'enic fluctuations from the solar/stellar convection zone to their dissipation in the wind \citep{suzuki2011self, cranmer2011testing, arber2016alfven}. However with the current advances in both, in-situ measurements of the solar wind, and remote-sensing observations from both the ground \citep{rimmele2020daniel} and the network of Heliospheric observers \citep{auchere2020coordination}, these analytic models are limited by the temporal scales and degree of non-linearity that they can capture. Thus codes like \bifrost{}, which are well suited to study small-scale dynamics, may be needed to feed back into the global wind models and analytic scaling relations.

\section{Conclusions}
In this study, the \bifrost{} RMHD code is used to produce a realistic simulation of the solar atmosphere. The simulation is configured to emulate a coronal hole region, with a uni-polar magnetic field in the low-corona that would later open into the solar wind. The photospheric magnetic field of the coronal hole is driven into Magnetic Concentrations (MCs) by the granulation motion. The movement of photospheric flux around the network of intergranular lanes effectively braids together the coronal magnetic field to form flux rope-like structures. If a MC enters the intersection of multiple granules, the strong field can suppress the local magnetoconvection and trigger the formation of a whirlpool-like flow which can drive the magnetic field to rotate and effectively \textit{stir} the low-corona.

Stirring motions are associated with an enhanced Poynting flux, which heats the plasma and drives material up into the low-corona. Vertical Poynting fluxes as large as $2-4$ kWm$^{-2}$, can persist for minutes at a time inside twisted magnetic funnel structures. The spatial distribution of heating of the low-corona is established by the underlying braided magnetic field.  The patchy-nature of this heating leads to structure at the base of the solar wind that is on the order of super-granulation. This is a similar scale to what is inferred for the modulation of velocity, temperature/density, and switchback occurrence in the recent observations from PSP \citep{fargette2021characteristic,bale2021solar}. 

The observational phenomena of chromospheric swirls (in quiet-Sun regions) may be an indicator of the same kind of stirring motions \citep[e.g.,][]{shetye2019multiwavelength}. The detection of chromospheric swirls in a coronal hole region would therefore be a useful diagnostic of the energy input to the solar wind. Stirring motions are a natural consequence of the interplay of convection and the network of magnetic funnels in coronal holes. Their potential ubiquity at the base of the solar wind may increase the frequency of switchback generation by Alfv\'enic turbulence in the solar wind (Shoda et al. in Prep). At present, the turbulent generation of switchbacks is unable to explain the occurrence rate of switchbacks observed by PSP \citep{shoda2020alfv}. The strength of the Alfv\'en wave-energy flux ($S_A/B=3.8\times 10^{5}$ Wm$^{-2}$T$^{-1}$) at the top of our simulation domain is weaker than required by current models of the solar wind that are driven by Alfv\'en wave-turbulence \citep[e.g.][]{shoda2019three}. However, due to the braiding of the coronal magnetic field, energy can be more rapidly channelled into the low-corona. This may off-set the decreased amplitude of fluctuations in the overall energy budget of the solar wind.

\subsection{Limitations of the Simulated Coronal Hole Patch} \label{limitations}

The solar wind is accelerated over a large domain, essentially from the solar surface (or {\it coronal base}) to a few solar radii, or some 10 coronal scale heights above the photosphere, where the wind becomes supersonic. The solar wind is primarly driven by the pressure gradient; as long as the corona is sufficiently hot and the magnetic field open to interplanetary space a supersonic wind will ensue \citep[see e.g][]{hansteen2012}. This heating can take the form of a Poynting flux carried by Alfv{\'e}n waves, braiding of field lines, or some other mechanism. In addition, in order to ensure a {\it fast} wind energy must be added beyond the critical point as pointed out by \citet{leer1980}, this additional push is often assumed to be caused by Alfv{\'e}n waves. Thus interpretation of our simulation results, in the context of the solar wind, must acknowledge that our domain only spans a third of a coronal scale height, and contains a horizontal size that is only slightly larger than a supergranule. 

As stated in the preceding sections a major stumbling block has been achieving a good understanding of how Alfv{\'e}nic waves are generated and carry a sufficient Poynting flux to push the wind and potentially heat the corona. The simulations presented here show that photospheric motions and their work on the magnetic field can drive the generation of Alfv{\'e}n waves of sufficient strength to be of interest to solar wind acceleration. This conclusion can potentially reduce the number of free parameters that go into (global) solar and stellar wind models. That said, it would of course be of interest to extend the 3D simulations to greater heights, at least one coronal scale height, and horizontally so that several supergranular cells can be accommodated. The latter would also require that the simulation extends a further distance below the photosphere, to at least of order $10$~Mm. 

Furthermore, it is important to note that the upper boundary in the simulations presented here are not perfectly transmissible to Alfv{\'e}n waves, and that reflections off the top boundary give rise to an inwardly propagating wave flux. Such an inwardly propagating flux is expected \citep[see][]{verdini2007} from models of Alfv{\'e}n wave dissipation in the solar wind where the outwardly propagating wave flux is reflected off perturbations in the solar wind's atmosphere, but perhaps not with the amplitude due to our imperfect open boundary. We therefore have not discussed the inwardly directed Alfv{\'e}n wave flux in any detail.

Running full 3D simulations spanning the upper convection zone through the photosphere and chromosophere to the first coronal scale height is a challenge; the scales are relatively large compared to the resolution required to accurately model the generation of Alfv{\'e}nic waves by photospheric convection or chromospheric dynamics and only a small sub-set of the configurations desired are possible to do in a finite time. This last point should be alleviated by the enormous technical progress seen in high performance computers in the last decades which should allow us to run a sufficient number of scenarios to complete parameter studies, both for solar and stellar configurations.

\begin{acknowledgements}
This research has received funding from the European Research Council (ERC) under the European Union’s Horizon 2020 research and innovation programme (grant agreement No 810218 WHOLESUN) and
by the Research Council of Norway through its Centres of Excellence scheme, project number 262622, and through grants of computing time from the Programme for Supercomputing. Some of the work was also performed with support from NASA's HTMS grant 80NSSC20K1272 ``Flux emergence and the structure, dynamics, and energetics of the solar atmosphere''.
We acknowledge funding support from INSU/PNST and CNES Solar Orbiter and Space Weather programs.
Data manipulation and Fourier Transforms were performed using the numpy \citep{2020NumPy-Array} and scipy \citep{2020SciPy-NMeth} python packages.
Figures in this work are produced using the python package matplotlib \citep{hunter2007matplotlib}.

\end{acknowledgements}

% WARNING
%-------------------------------------------------------------------
% Please note that we have included the references to the file aa.dem in
% order to compile it, but we ask you to:
%
% - use BibTeX with the regular commands:
%   \bibliographystyle{aa} % style aa.bst
%   \bibliography{Yourfile} % your references Yourfile.bib
%
% - join the .bib files when you upload your source files
%-------------------------------------------------------------------

\bibliographystyle{yahapj}
\bibliography{adam}

\begin{appendix}
\section{Developing Complexity From The Public Snapshots}\label{public}

In this work, the \simulation{} simulation is continued from a publicly available snapshot in order to allow magnetic complexity to further develop in the low-corona. As twisted coronal structures are primarily generated by the advection of MCs around the intergranular lanes, they take time to form in the domain. Firstly, the convection must expel the photospheric field into the intergranular lanes, forming concentrations in the magnetic field of sufficient strength to be braided together and produce long-lasting structure in the low-corona. Secondly, as the MCs are shuffled around the intergranular lanes at a speed of $\sim 5$~Mm/hour, this sets a characteristic timescale for this kind of structure to develop. As such, a few hours are required for the $24 \times 24$~Mm$^2$ domain to be sufficiently traversed by MCs, and reach a quasi-steady state. This is reflected in the temperature structure in the simulated atmosphere. 

Figure \ref{figureOldNew} shows four snapshots from the \simulation{} simulation, each taken one hour apart. These snapshots span from the start of the publicly available dataset, to the end of the dataset used in the study. During the course of the public dataset the temperature of the low-corona is observed to warm substantially from around 0.5~Mk to a more reasonable 0.9-1~Mk. The coronal temperature evolves slowly during this period as larger and larger Poynting fluxes are transmitted through the domain by braided magnetic field lines. The coronal magnetic field at the start of the computation strongly resembles the initial vertical magnetic field added to the \bifrost{} simulation. After an hour of computation, the simulation domain contains many twisted/braided magnetic field structures, and the magnetic field in the photosphere is more spatially-concentrated. The onset of widespread SEs in the simulation domain starts around $t=4,000$~s. As noted by \citet{2022arXiv220301221S}, who analysed a snapshot of \simulation{} at $t=4,610$~s. During the hour of simulation time investigated in this work ($t=9,580-13,180$~s), the temperature structure in the domain remains quasi-stationary (as shown in Figure \ref{figureAverageProfiles}). However some relaxation of the simulation is still expected, given the thermal relaxation timescale of the plasma in the computational domain.    

\begin{figure*}[t!]
 \centering
  \includegraphics[trim=0.0cm 0.0cm 0.0cm 0.0cm,clip, width=\textwidth]{a1.png}
   \caption{Comparison of the 3D magnetic field and temperature structure in the \simulation{} simulation domain in the publicly available dataset ($t=2,490-8,670$~s) and that used in this study ($t=9,580-13,180$~s). The four snapshots are each taken one hour apart (times are given with respect to the start of the hour used in this study). The magnetic field lines are coloured by plasma temperature. The photosphere is indicated in each case, and coloured by the vertical flow speed.}
   \label{figureOldNew}
\end{figure*}

\section{Reconstructed Frequency-filtered Velocity Components}\label{reconstruction}

\begin{figure*}[t!]
 \centering
  \includegraphics[trim=0.0cm 0.0cm 0.0cm 3.5cm,clip, width=\textwidth]{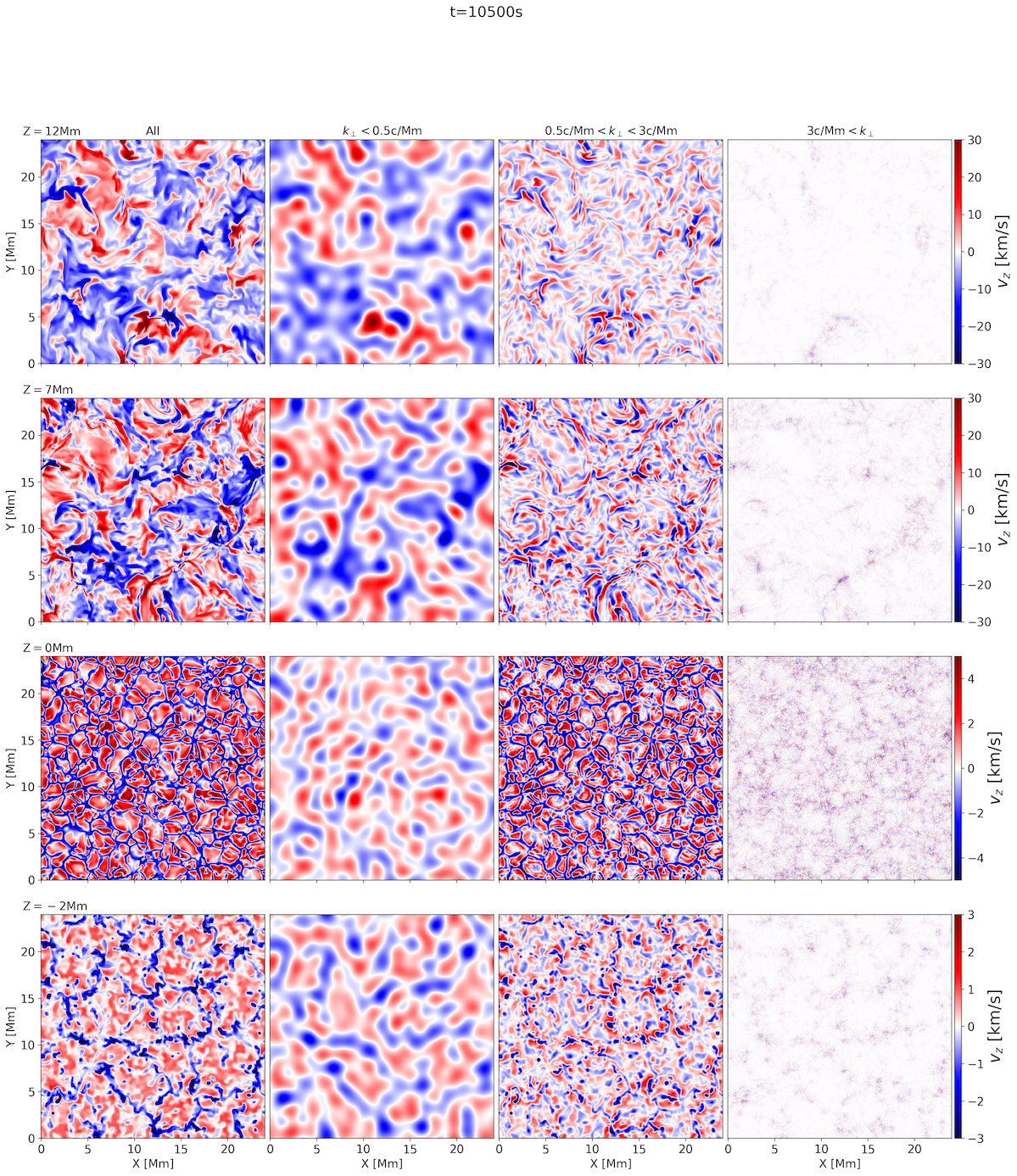}
   \caption{Vertical velocity reconstructed at $t=40$~mins by filtering the wavenumbers used in the inverse Fourier transform (the panels on the left show a slice through the simulation domain without any filtering).}
   \label{figureFilteringVertical2}
\end{figure*}

\begin{figure*}[t!]
 \centering
  \includegraphics[trim=0.0cm 0.0cm 0.0cm 3.5cm,clip, width=\textwidth]{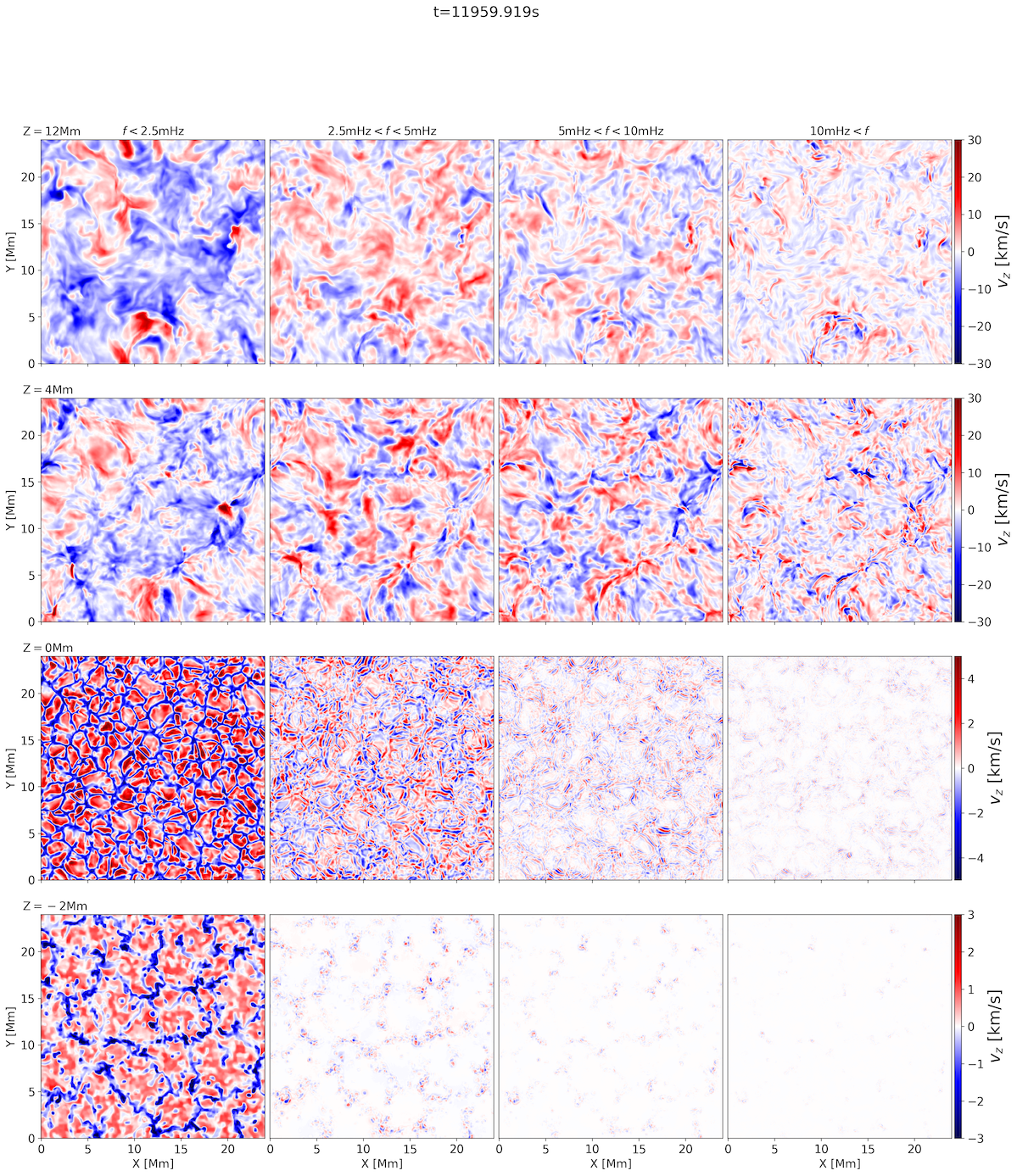}
   \caption{Vertical velocity reconstructed at $t=40$~mins by filtering the frequencies used in the inverse Fourier transform.}
   \label{figureFilteringVertical}
\end{figure*}

\begin{figure*}[t!]
 \centering
  \includegraphics[trim=0.0cm 0.0cm 0.0cm 3.5cm,clip, width=\textwidth]{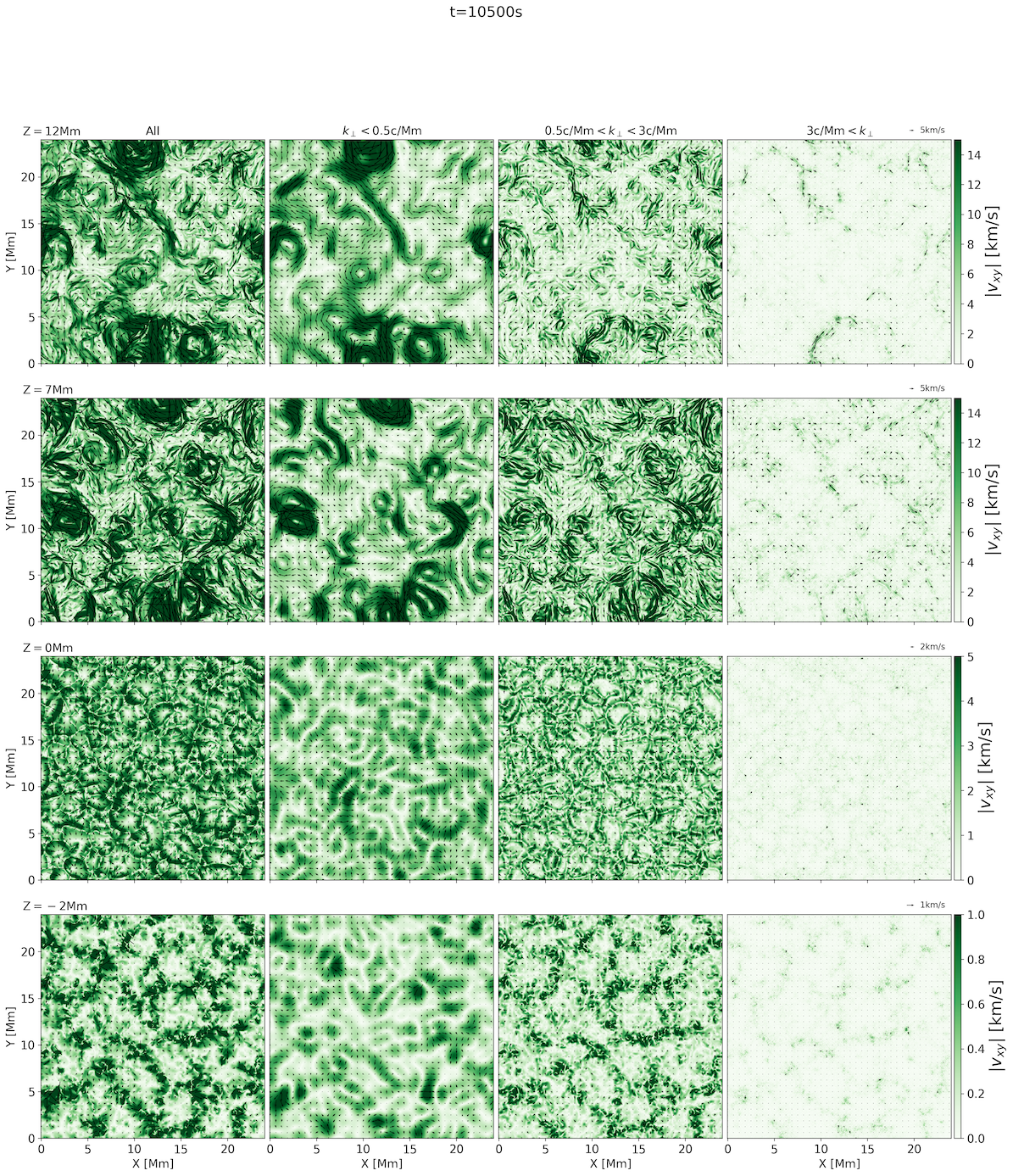}
   \caption{Same as Figure \ref{figureFilteringVertical2}, but now for the magnitude of the horizontal flow speed $|v_{xy}|$. Arrows indicate the direction of the flow in the x-y plane. }
   \label{figureFilteringHorizontal2}
\end{figure*}

\begin{figure*}[t!]
 \centering
  \includegraphics[trim=0.0cm 0.0cm 0.0cm 3.5cm,clip, width=\textwidth]{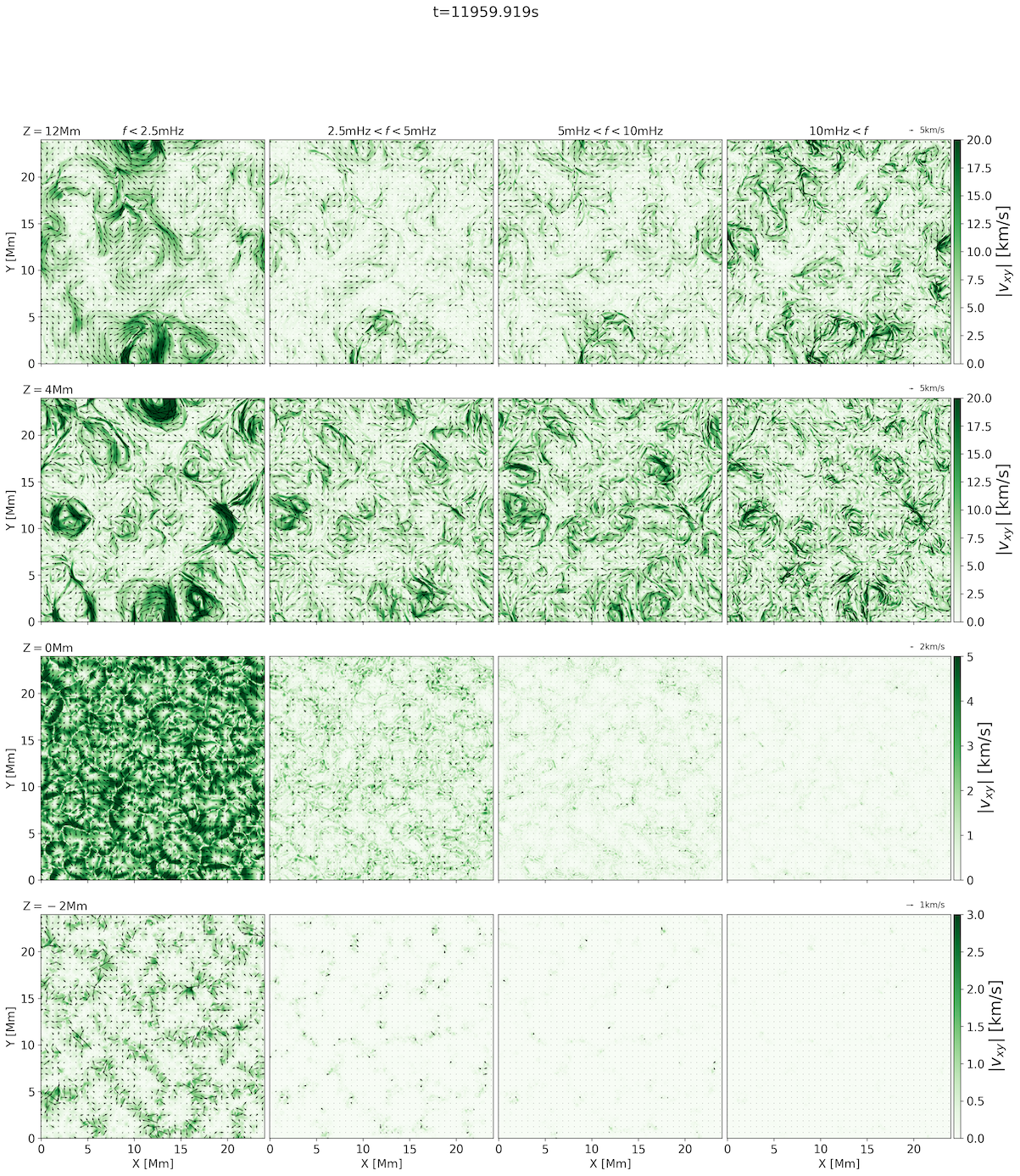}
   \caption{Same as Figure \ref{figureFilteringVertical}, but now for the magnitude of the horizontal flow speed $|v_{xy}|$. }
   \label{figureFilteringHorizontal}
\end{figure*}

From the Fourier transformations $F_i$ of the velocity components $v_z$ and $v_{xy}$ in Section \ref{waves}, it is possible to reconstruct the strength of each component in selected wavenumber/frequency ranges by performing the inverse Fourier transform and using only the Fourier coefficients from those chosen frequencies. For the same four heights shown in Figures \ref{figureVzFFT} and \ref{figureVxyFFT}, the flow speeds in three wavenumber ranges: less than 0.5 cycles/Mm, 0.5 - 3 cycles/Mm and greater than 3 cycles/Mm, along with four frequency ranges: less than 2.5 mHz, 2.5 - 5 mHz, 5 - 10 mHz, and greater than 10 mHz, are recovered. The reconstructed velocity components at $t=40$~mins are displayed in Figures \ref{figureFilteringVertical2}, \ref{figureFilteringVertical},  \ref{figureFilteringHorizontal2} and \ref{figureFilteringHorizontal}. For the horizontal velocity, the colour displays the magnitude of the flow speed and the arrows indicate the flow direction. In the upper-convection zone, the flow is dominated by the slowly evolving convective motions (as shown in Figures \ref{figureVzFFT} and \ref{figureVxyFFT}). At the photosphere, these slow vertical convective motions become the recognisable granulation pattern in the lowest frequency range, and the Mm-scale wavelength range. The circulation of plasma from the up-welling of hot material in the center of the granule to the cooler down-flow lanes generates a low frequency horizontal flow (including vortical flows around complex intersections). The forcing from convective granules in the photosphere generates wave-trains, that are visible in the two middle frequency ranges of $v_z$ in Figure \ref{figureFilteringVertical} at Z = 0 Mm. These waves experience less resistance to their propagation in the intergranular lanes. They may also gain energy from the box-modes that resonate inside the convection zone. As the highest frequency waves are mostly confined to the intergranular lanes at the photosphere. High frequency fluctuations can influence the plasma above by interacting with the MCs that also reside in the intergranular lanes, and extend up into the low-corona. 

In the low-corona, the low frequency/wavenumber motions are dominated by stirring from the magnetic funnels and SEs. The vertical components shows the hot rising material associated with the heating from twisted magnetic field structures, and cool sinking material elsewhere. The horizontal flow is organised into clear swirls which follow the twisted magnetic field structures. The higher frequency components have increased in amplitude through the chromosphere/transition region. Theses frequencies are dominated by torsional oscillations inside the twisted magnetic field structures. Towards the top of the domain, the strength of the horizontal component in the two middle frequency ranges begins to dissipate (also visible in Figure \ref{figureVxyFFT}), which is not observed for the highest frequencies. Comparing the PSD at 4 Mm and 12 Mm in Figures \ref{figureVzFFT} and \ref{figureVxyFFT}, there appears to be little change in the low-corona for vertical fluctuations. Waves mostly propagate towards the top of the domain and escape through the open boundary condition. The dominant frequencies in the low-corona for both the velocity components correspond to the vertical convective motions, and the box-modes.

\end{appendix}

\end{document}